\documentclass{pasa}
\usepackage{longtable}

\title[Long-term multicolor photometry of the YSOs]{Long-term multicolor photometry of the young stellar objects FHO 26, FHO 27, FHO 28, FHO 29 and V1929 Cygni}
\author[Ibryamov et al.]{S. I. Ibryamov$^{1}$\thanks{E-mail: sibryamov@astro.bas.bg}, E. H. Semkov$^1$, S. P. Peneva$^1$\\
\affil{$^1$Institute of Astronomy and National Astronomical Observatory, Bulgarian Academy of Sciences,
              72, Tsarigradsko Shose Blvd., BG-1784 Sofia, Bulgaria}}%
\jid{PASA}
\doi{10.1017/pas.\the\year.xxx}
\jyear{\the\year}

\begin{document}

\begin{abstract}
Results from long-term multicolor optical photometric observations of the pre-main sequence stars FHO 26, FHO 27, FHO 28, FHO 29 and V1929 Cyg collected during the period from June 1997 to December 2014 are presented. The objects are located in the dense molecular cloud L935, named ''Gulf of Mexico'', in the field between the North America and Pelican nebulae. All stars from our study exhibit strong photometric variability in all optical passbands. Using our $BVRI$ observations and data published by other authors, we tried to define the reasons for the observed brightness variations. The presented paper is a part of our long-term photometric study of the young stellar objects in the region of ''Gulf of Mexico''.

\end{abstract}
\maketitle%
\section{INTRODUCTION}
\label{sec:intro}


Pre-main sequence (PMS) stars are newly formed objects located above the main sequence on the Hertzsprung-Russell diagram. 
Depending on their initial mass these stars reach the main sequence for different time scales. 
One of the most important features of the PMS stars is their photometric variability.
PMS stars are important objects for study of the early stages of the stellar evolution and for testing the stellar evolution scenarios.

PMS stars with low masses (M $\leq$ 2M$_{\odot}$) are named T Tauri stars (TTS). 
The first systematic study of such young stellar objects, as a separate class with prototype the star T Tauri, was made by Joy \shortcite{joy, jah}. 
TTS are associated with obscured regions, molecular clouds and dark nebulae and are grouped in T associations. 
These objects show strong irregular photometric variability and emission line spectra. 
TTS are separated into two subclasses $-$ Classical T Tauri stars (CTTS), which are surrounded by massive accreting circumstellar disks, and weak-line T Tauri stars (WTTS) without evidence for disk accretion \cite{ber, men}. 

The more massive (2M$_{\odot}$ $\leq$ M $\leq$ 8M$_{\odot}$) PMS stars are known as Herbig Ae/Be stars (HAEBES). 
The two types PMS stars $-$ TTS and HAEBES are still contracting towards the main sequence and their main characteristic is the photometric and spectroscopic variability, discovered at the beginning of their study.

TTS show various types of stellar activity. According to Herbst et~al. \shortcite{hem} the variability of WTTS is due to cool spots or groups of spots on the stellar surface. 
The photometric amplitudes of this variability are about 0.03-0.3 mag, but in extreme cases reach 0.8 mag in the $V$-band.  
The variability of WTTS can also be due to flare-like variations in $B$- and $U$-band. Flares are random with different sized amplitudes, as there is no periodicity. Usually, the rise of brightness is quick and occurs over short time scales. CTTS often shows irregular variations with larger photometric amplitudes up to 2-3 mag, associated with high variable accretion from the circumstellar disk onto the stellar surface. The variability of CTTS also can be due to rotating hot spots on the stellar surface.

The large amplitude drops in the brightness of the PMS stars are commonly observed in the early types of TTS and HAEBES. 
These objects show non-periodic deep Algol-type minima and amplitudes up to 2.8 mag in $V$-band. 
The observed drops in the brightness last from days to some weeks and are caused by circumstellar dust or clouds obscuration \cite{vos, gri, hem}. 
This group of PMS objects with intermediate mass is known as UXors and their prototype is the star UX Orionis. 
In very deep minima the color indixes of UXors often becomes bluer (so called ''blueing effect'' or ''color reverse'') (see Bibo \& Th\'{e} \shortcite{bib}). 
It is generally accepted that the origin of observed drops in the brightness and the bluing effect are due to variations of the column density of dust in the line of sight to the star \cite{dul}.

The outbursts of the PMS stars with large amplitude are grouped into two types $-$ FUors, with prototype the star FU Orionis, and EXors, with prototype the star EX Lupi \cite{hgh}.
During the quiescence state the FUors and EXors are normally accreting CTTS. 
The outbursts of both types of eruptive PMS stars are generally attributed to a sizable increase in accretion rate from the circumstellar disk onto the stellar surface.

The PMS stars included in the present study are located in the dense molecular cloud L935, known as ''Gulf of Mexico'', between North America (NGC 7000) and Pelican Nebula (IC 5070), in the vicinity of the new FUor star V2493 Cygni (HBC 722) erupted in 2010 \cite{sem, spm, spi, mil}. 
NGC 7000 and IC 5070 are thought to be parts of a single large HII region W80. 
The distance to this region, determined by Laugalys \& Strai\v{z}ys \shortcite{lau} is 600 kpc. 
''Gulf of Mexico'' is a region with active star formation and contains many young stellar objects (YSO) \cite{arm, fin, bal}.
The stars FHO 26, FHO 27, FHO 28 and FHO 29 are revealed as large amplitude variables in the study of Findeisen et al. \shortcite{fin}.
V1929 Cyg is reported as flare star from UV Ceti type by Rosino et al. \shortcite{ros}.

The present paper is a part of our long-term photometric study of the PMS stars in the region of ''Gulf of Mexico''. 
The results from our previous studies of PMS stars in this field have been published in Semkov et al. \shortcite{sem}, \shortcite{spm}, \shortcite{spi}; Poljan\v{c}i\'{c} Beljan et al. \shortcite{pjs}; Ibryamov et al. \shortcite{ibr}.

\section{OBSERVATIONS AND DATA REDUCTION}

The presented data from photometric observations were performed during the period from June 1997 to December 2014, in two astronomical observatories, with four telescopes. These are the 2-m Ritchey-Chr\'{e}tien-Coud\'{e} (RCC), the 50/70-cm Schmidt and the 60-cm Cassegrain telescopes of the Rozhen National Astronomical Observatory of the Bulgarian Academy of Sciences and the 1.3-m Ritchey-Chr\'{e}tien (RC) telescope of the Skinakas Observatory\footnote{Skinakas Observatory is a collaborative project of the University of Crete, the Foundation for Research and Technology, Greece, and the Max-Planck-Institut f{\"u}r Extraterrestrische Physik, Germany.} of the University of Crete (Greece).

The observations were performed with eight different types of CCD cameras Photometrics AT200 and VersArray 1300B at the 2-m RCC telescope; Photometrics CH360 and ANDOR DZ436-BV at the 1.3-m RC telescope; SBIG ST-8, SBIG STL-11000M and FLI PL16803 at the 50/70-cm Schmidt telescope; FLI PL09000 at the 60-cm Cassegrain telescope. The technical parameters and specifications of the used CCD are summarized in Table~\ref{Tab1}. 

\begin{table*}
\caption{CCD cameras used and their technical parameters and specifications} 
\begin{center}
\begin{tabular*}{\textwidth}{@{}l\x l\x c\x c\x c\x c\x c\x c\x @{}}
\hline \hline
Telescope & CCD Camera & Chip size & Field & Pixel size & Scale & RON & Gain\\
          &            & [\textit{pix}$\times$\textit{pix}] & [$arcmin$ $\times$ $arcmin$] & [$\mu m$] & [$arcsec$/\textit{pix}] & [\textit{e$^{-}$ rms}] & [\textit{e$^{-}$/ADU}]\\
\hline 
2-m RCC    & Photometrics AT200  & 1024 $\times$ 1024 & 5.28 $\times$ 5.28 & 24 & 0.31 & 3.90 & 4.9\\
2-m RCC    & VersArray 1300B     & 1340 $\times$ 1300 & 5.76 $\times$ 5.59 & 20 & 0.26 & 2.00 & 1.0\\
\hline
1.3-m RC   & Photometrics CH360  & 1024 $\times$ 1024 & 8.57 $\times$ 8.57 & 24 & 0.50 & 2.60 & 2.3\\
1.3-m RC   & ANDOR DZ436-BV      & 2048 $\times$ 2048 & 9.64 $\times$ 9.64 & 13.5& 0.28 & 8.14 & 2.7\\
\hline
Schmidt    & SBIG ST-8           & 1530 $\times$ 1020 & 27.52 $\times$ 18.35 & 9 & 1.08 & 6.20 & 2.3\\
Schmidt    & SBIG STL-11000M     & 4008 $\times$ 2672 & 72.00 $\times$ 48.00 & 9 & 1.08 & 13.00 & 0.8\\
Schmidt    & FLI PL16803         & 4096 $\times$ 4096 & 73.80 $\times$ 73.80 & 9 & 1.08 & 9.00 & 1.0\\
\hline
Cassegrain & FLI PL09000         & 3056 $\times$ 3056 & 16.81 $\times$ 16.81 & 12 & 0.33 & 8.50 & 1.0\\
Cassegrain$^*$ & FLI PL09000     & 3056 $\times$ 3056 & 26.40 $\times$ 26.40 & 12 & 0.52 & 8.50 & 1.0\\
\hline \hline
\end{tabular*}\label{Tab1}
\end{center}
\tabnote{$^*$ with focal reducer}
\end{table*}

The photometric limit of our data obtained with the 2-m RCC and the 1.3-m RC telescopes in $B$- and $V$-band is about 20.5 mag and for the data collected with the 50/70-cm Schmidt and the 60-cm Cassegrain telescopes is about 19.5 mag.

All frames were taken through a standard Johnson-Cousins set of filters. The total number of the nights used for observations is 292. All frames obtained with the VersArray 1300B, Photometrics CH360 and ANDOR DZ436-BV cameras are bias frame subtracted and flat field corrected. The frames obtained with the SBIG ST-8, SBIG STL-11000M, FLI PL16803 and FLI PL9000 cameras are dark frame subtracted and flat field corrected.

The photometric data were reduced by IDL package (standard subroutine $DAOPHOT$). As a reference, the $BVRI$ comparison sequence reported in Semkov et.~al. \shortcite{sem} was used. All data were analyzed using the same aperture, which was chosen to have a 4 arcsec radius, while the background annulus was taken from 9 arcsec to 14 arcsec. 

The average value of the errors in the reported magnitudes are 0.01$-$0.05 mag for $I$- and $R$-band data, 0.02$-$0.08 mag for $V$-band data and 0.03$-$0.11 mag for $B$-band data.

\section{RESULTS AND DISCUSSION}

\begin{figure}
\begin{center}
\includegraphics[width=\columnwidth]{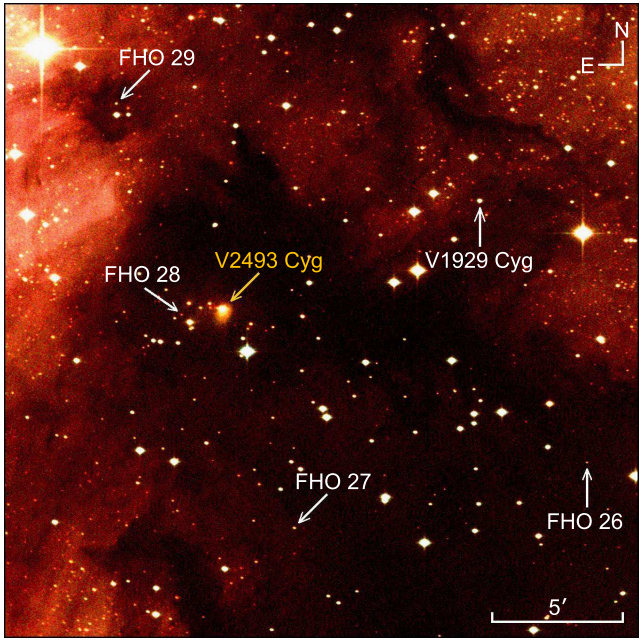}
\caption{An three-color image of the field of ''Gulf of Mexico'', obtained with the Schmidt telescope of Rozhen NAO on August 19, 2014. The program stars and the recently erupting FUor star V2493 Cyg are marked.}
\label{Fig1}
\end{center}
\end{figure}

The results from long-time $BVRI$ photometric observations of the PMS stars FHO 26, FHO 27, FHO 28, FHO 29 and V1929 Cyg are presented in the paper. 
Figure~\ref{Fig1} shows an three-color image of the field of ''Gulf of Mexico'', with marked positions of the stars from our study and the recently erupting FUor star V2493 Cyg. The image was obtained on August 19, 2014 with the 50/70-cm Schmidt telescope of Rozhen NAO.

The stars from our study shows bursts and/or fades events in their light curves. These events are defined as follows. In the case that the star most of the time spends at high light during our study, the declines in the brightness are interpreted as fades. Respectively, if the star most of the time spends at a low light, the observed increases in the brightness are interpreted as bursts.

\subsection{FHO 26}

The star FHO 26 (2MASS J20565934+4347530) is included in the list of YSO candidates published in Guieu et~al. \shortcite{gui}. 
The authors measured $B$ = 20.195, $V$ = 18.747 and $I$ = 15.963 magnitudes of the star. According to Findeisen et~al. \shortcite{fin} FHO 26 shows bursts events, which are repeated every few weeks.

Findeisen et~al. \shortcite{fin} presented the light curve in $R$-band of FHO 26 for the period 2009$-$2012 and described $\sim$0.7 mag burst events in 2010 and 2011, lasting 4-5 days each and separated by 10-30 days. 
During 2012 the star has shown only two short bursts. 
The spectrum of FHO 26 taken in July 2012 by Findeisen et~al. \shortcite{fin} shows a M4.5 photosphere with H$\alpha$ in emission.

\begin{figure*}
\begin{center}
\includegraphics[width=\textwidth]{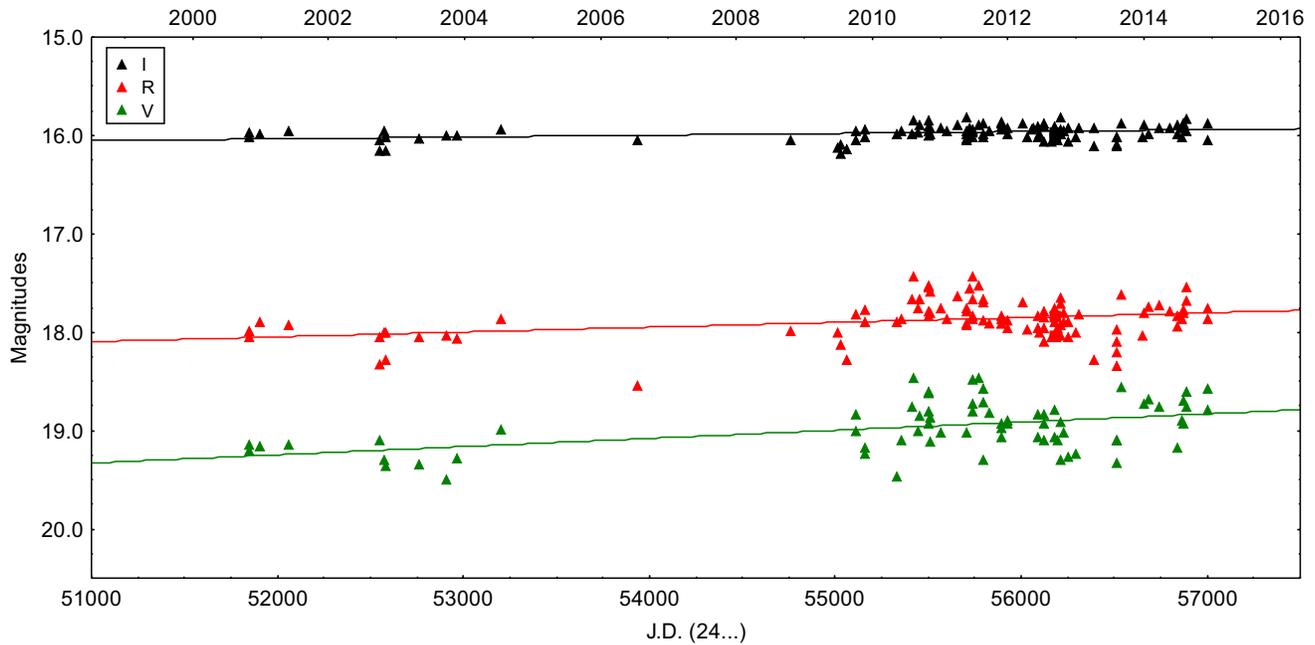}
\caption{CCD $IRV$ light curves of FHO 26 for the period October 2000$-$December 2014}
\label{Fig2}
\end{center}
\end{figure*}

The $VRI$ light curves of FHO 26 from our CCD observations are shown in Fig~\ref{Fig2}. On the figure, triangles denote CCD photometric data obtained with the 50/70-cm Schmidt telescope and squares $-$ the photometric data collected with the 60-cm Cassegrain telescope. The results of our long-term multicolor CCD observations of the star are summarized in Table~\ref{Tab2}. The columns contains date (DD/MM/YYYY format) and Julian data (J.D.) of the observations, $IRV$ magnitudes of the star, telescope and CCD camera used.

It can be seen from Figure~\ref{Fig2} that during our study the brightness of FHO 26 vary around some intermediate level. The star shows both increases in the brightness with different amplitudes and short drops in the brightness.
The brightness of the star during the period 2000$-$2014 vary in the range 15.81$-$16.18 mag for $I$-band, 17.43$-$18.54 mag for $R$-band and 18.47$-$19.50 mag for $V$-band. 
In a very low light the brightness of FHO 26 in $V$-band is under the photometric limit of the 50/70-cm Schmidt and the 60-cm Cassegrain telescopes. 
The observed amplitudes are $\Delta I$ $\sim$ 0.40 mag, $\Delta R$ $\sim$ 1.10 mag and $\Delta V$ $>$ 1.0 mag. 

The star shows a large amplitude photometric variability during certain periods. The observed increases of the star's brightness in the period August 2010$-$ January 2011 are with amplitudes up to 0.40 mag for $R$-band and up to 0.65 mag for $V$-band; in the period June$-$September 2011 with amplitudes up to 0.50 for $R$-band and up to 0.80 mag for $V$-band; in the period July 2012$-$January 2013 with amplitudes up to 0.40 mag for $R$-band and up to 0.50 mag for $V$-band and in the period June$-$August 2014 are with amplitudes up to 0.35 for $R$-band and up to 0.40 mag for $V$-band. Usually, variability with such amplitudes is an indication for the presence of hot spots on the stellar surface. 
The variability is produced by variable accretion from the circumstellar disk and it is typical of CTTS.

In October 2002, July 2006, August 2009, April 2013 and August 2013 small amplitude drops in the brightness of FHO 26 are observed. 
The amplitudes are small ($\Delta R$ $\sim$ 0.40 mag) and the drops are caused probably by the existence of cool spots on the stellar surface or by small irregular obscuration of the star by circumstellar material. It can be seen from Fig.~\ref{Fig2} that during the whole period of observations the total star's brightness gradually increases (well seen in $R$- and $V$-bands). 
The rates of increase are $\sim$0.69$\times$$10^{-2}$ mag $yr^{-1}$ for $I$-band, $\sim$1.79$\times$$10^{-2}$ mag $yr^{-1}$ for $R$-band and $\sim$3.04$\times$$10^{-2}$ mag $yr^{-1}$ for $V$-band

\begin{figure}
\begin{center}
\includegraphics[width=4cm]{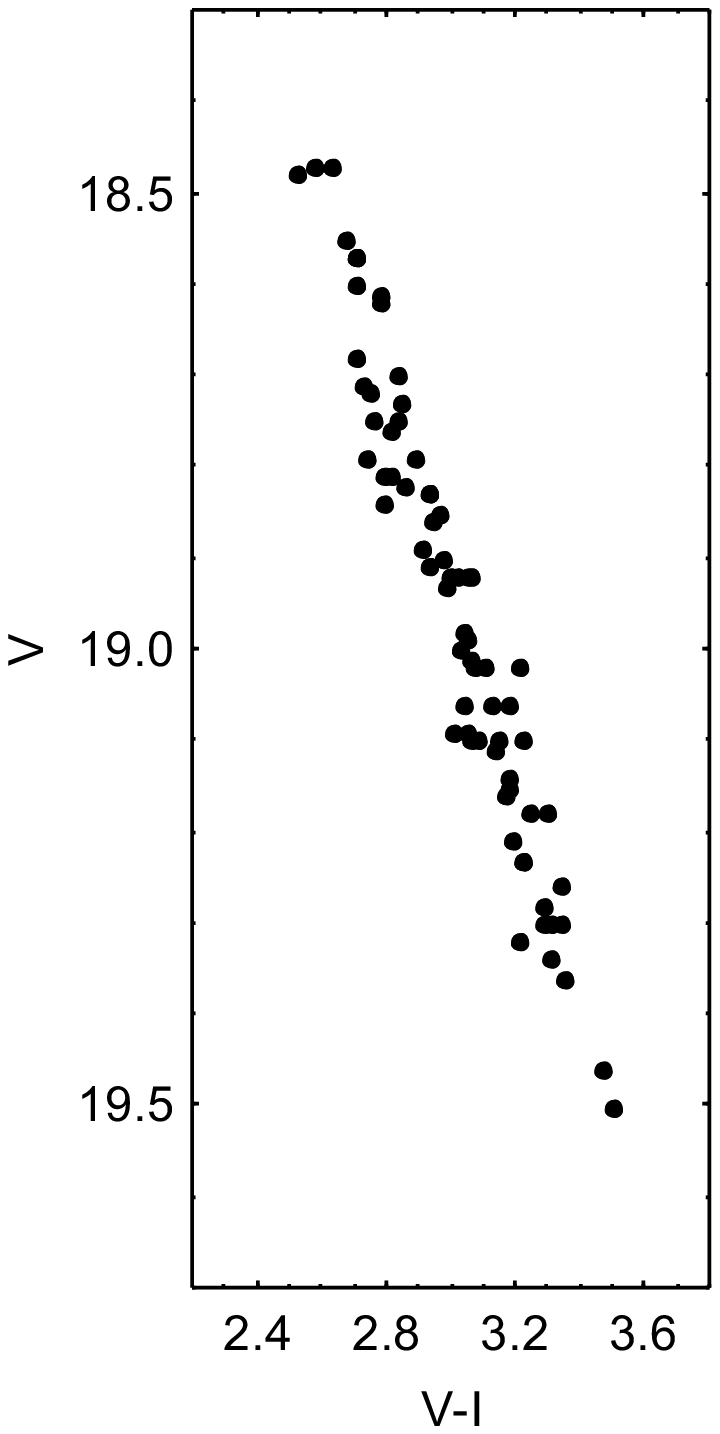}
\includegraphics[width=4cm]{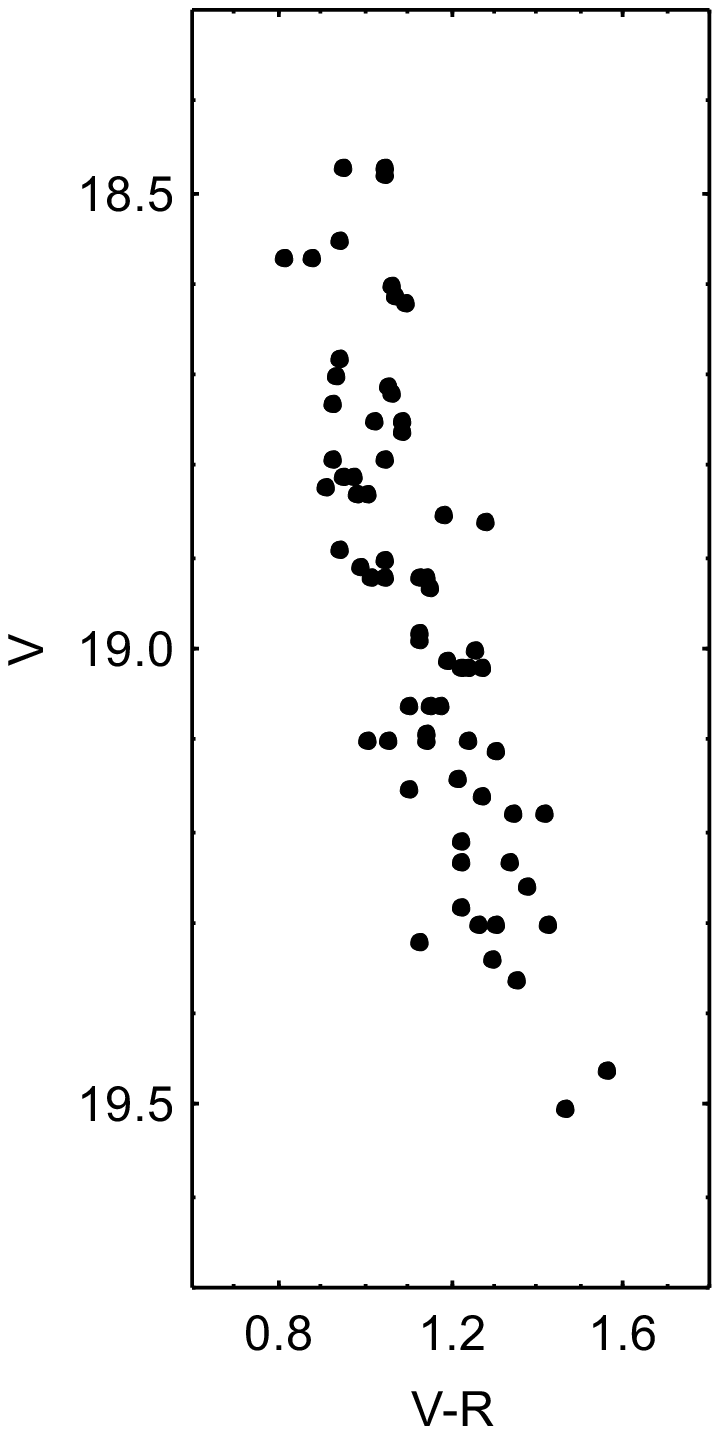}
\caption{Relationship between $V$ magnitude and $V - I$ and $V - R$ color indexes of FHO 26 in the period October 2000$-$December 2014}\label{Fig3}
\end{center}
\end{figure}

The measured color indexes $V - I$ and $V - R$ versus stellar $V$ magnitude for the period of our observations are plotted on Fig.~\ref{Fig3}. From the figure it is seen that the star becomes redder as it fades and color reverse was not observed. 
Such color variations are typical for both WTTS and CTTS, stars with cool spots, whose variability is produced by rotation of the spotted surface.
In case that the decreases in the brightness are caused by obscuration from circumstellar material, the observed amplitudes are too small to show indication for color reverse.

Therefore, the long-term light curves of FHO 26 gives grounds to predict different reasons for observed variability of the star. 
The light curves of FHO 26 shows photometric characteristics of both WTTS and CTTS with a presence of hot and cool spots on the stellar surface. 
Evidences of periodicity in the brightness variability are not detected. 
The spectrum with H$\alpha$ emission of FHO 26 \cite{fin} can be classified as a TTS spectrum.

\subsection{FHO 27}

The star FHO 27 (2MASS J20580138+4345201) is included in the list of YSO candidates by Guieu et~al. \shortcite{gui}. 
The authors measured $B$ = 20.001, $V$ = 18.319 and $I$ = 15.759 magnitudes of the star. Rebull et~al. \shortcite{rlm} classify FHO 27 as a flat spectrum source.

Findeisen et~al. \shortcite{fin} presented the light curve in $R$-band of the star for the period 2009$-$2012. According to the authors FHO 27 exhibits a non-periodic multiple fading events, lasting 15-40 days and separated by 30-60 days, as fading events get shallower over the course of the decline. 
The authors pay attention that FHO 27 was not detected in roughly half the epochs in 2012.

The spectrum of FHO 27 taken in July 2012 by Findeisen et~al. \shortcite{fin} during the stars long-term low state is typical of K7 photosphere with strong H$\alpha$, Paschen series, and CaII emission and weaker HeI, [OI] and OI lines.

\begin{figure*}
\begin{center}
\includegraphics[width=\textwidth]{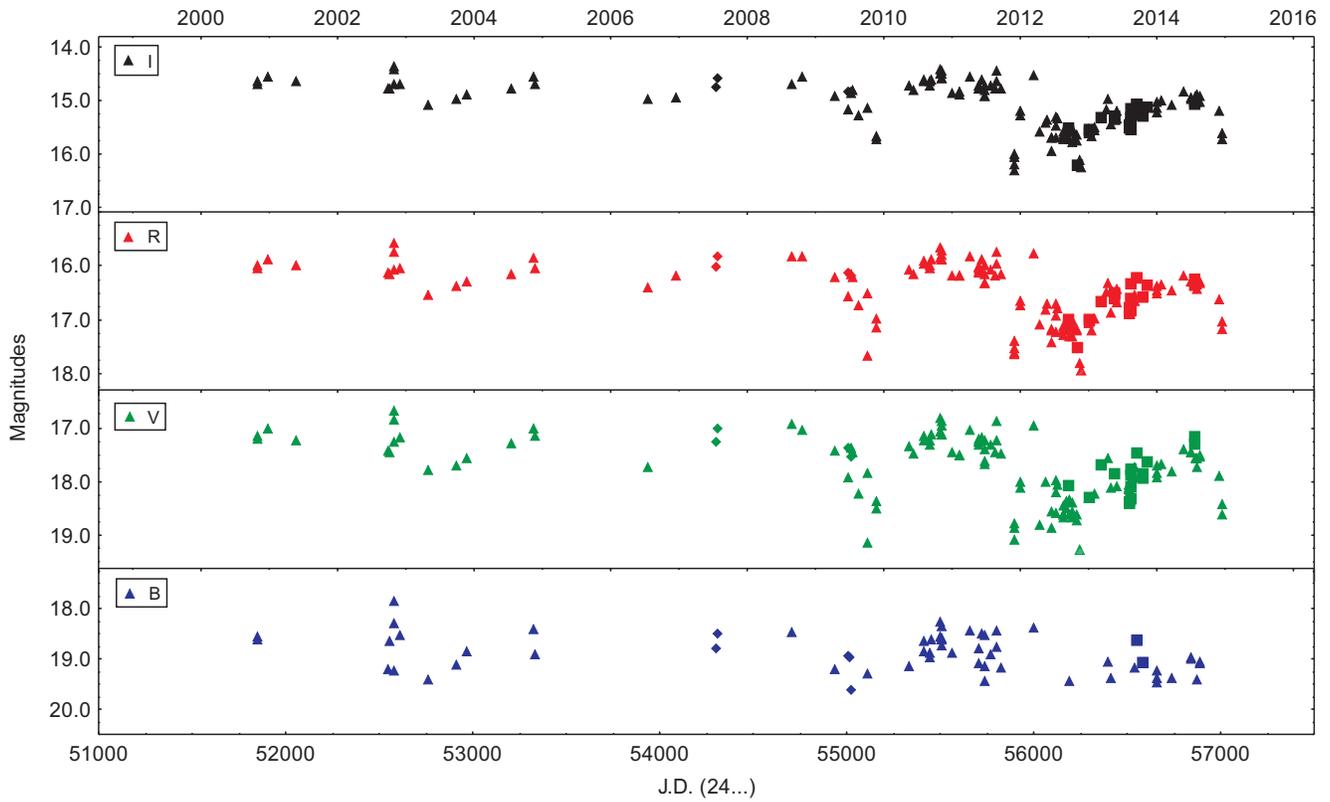}
\caption{CCD $IRVB$ light curves of FHO 27 for the period October 2000$-$December 2014}\label{Fig4}
\end{center}
\end{figure*}

The $BVRI$ light curves of FHO 27 from our CCD observations are shown in Fig.~\ref{Fig4}. In the figure, diamond denote CCD photometric data obtained with the 1.3-m RC telescope, the symbols used for the 50/70-cm Schmidt and the 60-cm Cassegrain telescopes, are as in Fig.~\ref{Fig2}. The results of our long-term multicolor CCD observations of FHO 27 are summarized in Table~\ref{Tab3}. The columns contains date (DD/MM/YYYY format) and Julian data (J.D.) of the observations, $IRVB$ magnitudes of the star, telescope and CCD camera used.

The brightness of FHO 27 during the period of our observations 2000$-$2014 vary in the range 14.36$-$16.31 mag for $I$-band, 15.58$-$17.93 mag for $R$-band, 16.68$-$19.26 mag for $V$-band and 17.84$-$19.62 mag in $B$-band. Because of the limit of our photometric data the deeper fading events of FHO 27 in $V$- and $B$-bands are not registered. 
The observed amplitudes are $\Delta I$ $\sim$ 1.95 mag, $\Delta R$ $\sim$ 2.35 mag, $\Delta V$ $>$ 2.6 mag and $\Delta B$ $>$ 1.8 mag.

It can be seen from Fig.~\ref{Fig4} that the star usually spends most of the time at high light. During our photometric monitoring several deep declines in the brightness of FHO 27 are registered. The drops in the brightness are non-periodic and have different duration and amplitudes. 
It is possible to suggest the existence of a frequent deep fading events during periods with insufficient data.
Now we can identify four deep fading events, at the end of 2009$-$beginning 2010, at the end of 2011, in the period 2012-2013, and in the period 2013$-$2014.

In the early 2012 the deepest and prolonged drop in the brightness in all bands of FHO 27 begins and reaches the biggest depth in November 2012 with amplitudes 1.82 mag in $I$-band, 2.20 mag in $R$-band and 2.38 mag for $V$-band. 
Since November 2012 the star's brightness began to rise, continuing to late 2014, but from November 2014 new drop in the brightness started. 
During the prolonged fading event some low amplitude sudden declines are also observed. Evidences of periodicity in the brightness variability of FHO 27 are not detected.

\begin{figure*}
\begin{center}
\includegraphics[width=4cm]{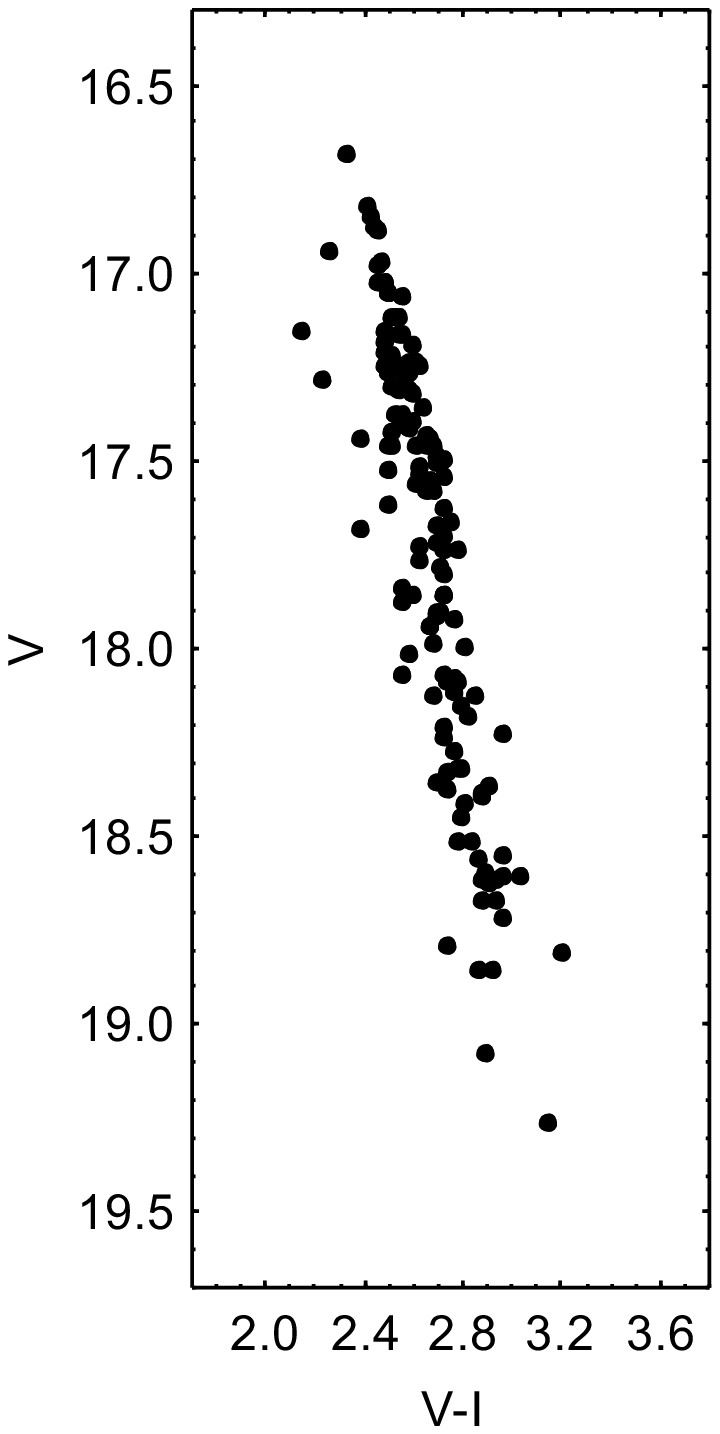}
\includegraphics[width=4cm]{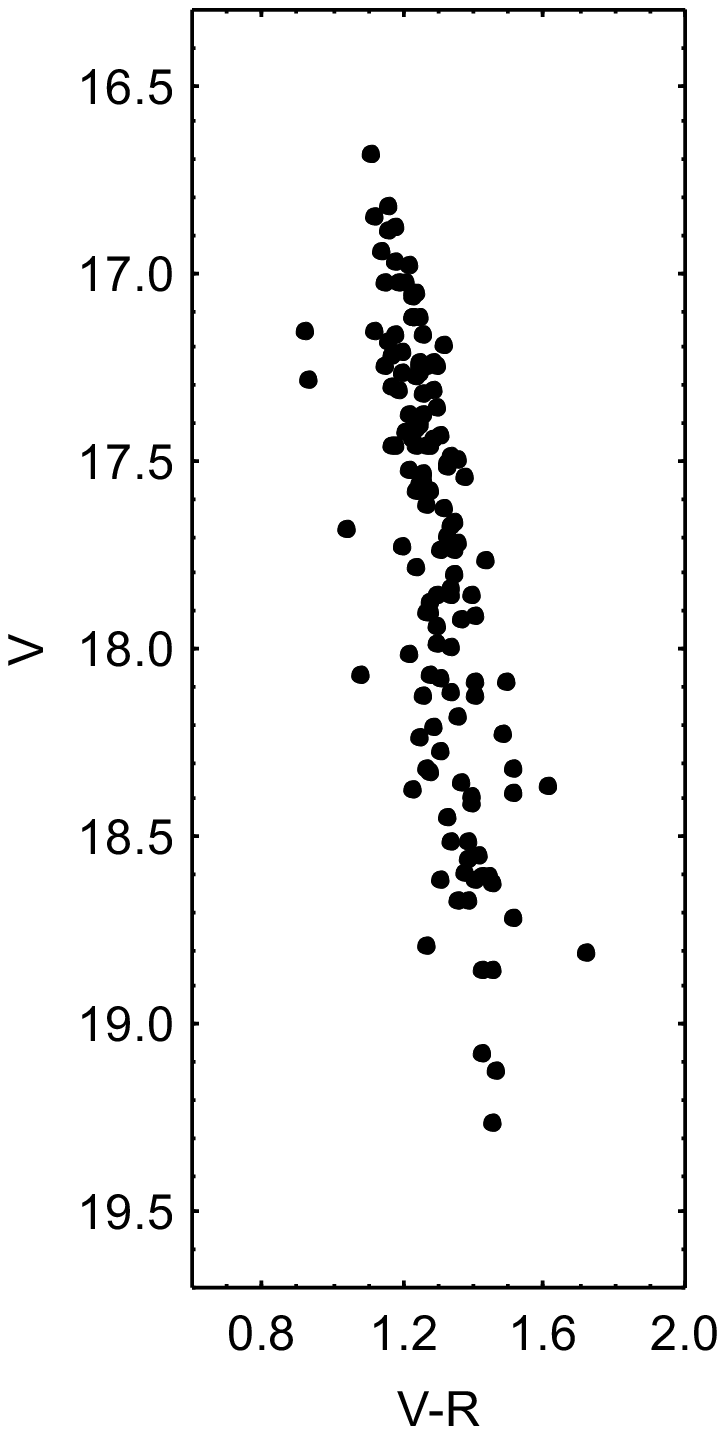}
\includegraphics[width=4cm]{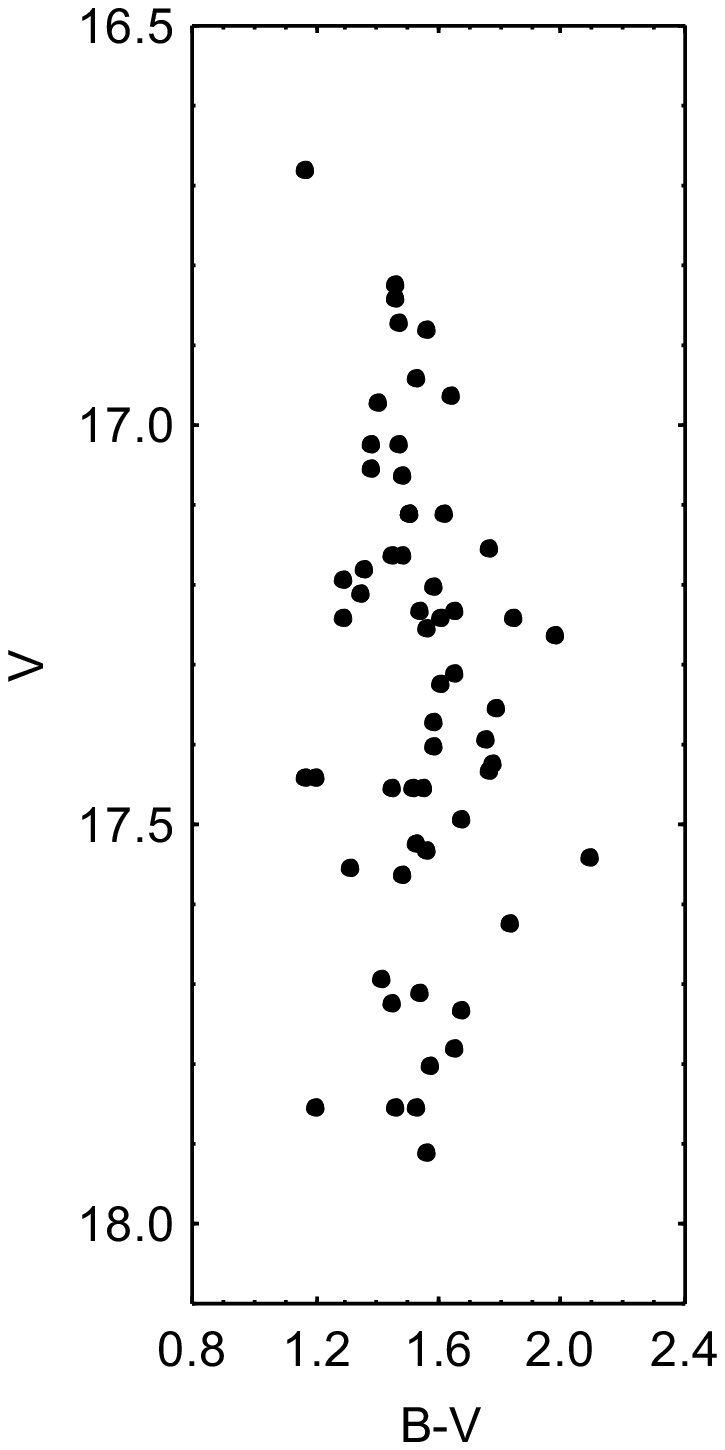}
\caption{Relationship between $V$ magnitude and the $V - I$, $V - R$ and $B - V$ color indexes of FHO 27 in the period October 2000$-$December 2014}\label{Fig5}
\end{center}
\end{figure*}

The large amplitudes of observed declines in the brightness of FHO 27 are indication of UXor-type variability and the deep fading events presumably are result of circumstellar dust or clouds obscuration. 
On Figure~\ref{Fig5} the color indexes $V - I$, $V - R$ and $B - V$ versus stellar $V$ magnitude during the period of our observations are plotted.
The figure shows evidences for color reversal, especially for $B - V$ index. 
The different shape of the observed declines gives grounds to predict a variety of obscuration reasons, clouds of proto stellar material, massive dust clumps orbiting the star or planetesimals at different stages of formation.

The light curves of FHO 27 are very similar to light curves of V521 Cyg (HBC 299), which was classified as UXor by Grankin et al. \shortcite{gra} and Ibryamov et al. \shortcite{ibr}. V521 Cyg is located at about 9 arcmin from FHO 27.

\subsection{FHO 28}

The star FHO 28 (2MASS J20582555+4353287) is included in the list of YSO candidates published in Guieu et~al. \shortcite{gui}. 
The authors measured $B$ = 21.012, $V$ = 19.153 and $I$ = 15.932 magnitudes of the star. 
Armond et~al. \shortcite{arm} includes FHO 28 in the list of H$\alpha$ emission-line stars and measured $V$ = 19.58, $R$ = 18.11 and $I$ = 16.81 magnitudes of the object. 

Findeisen et~al. \shortcite{fin} presented the light curve in $R$-band of FHO 28 for the period 2009$-$2012 and described fades events of the star. The spectrum of FHO 28 taken in 2012 by Findeisen et~al. \shortcite{fin} during the star's strongly varying phase shows a M3 photosphere with H$\alpha$ in emission and weak emission of CaII, [NII], HeI.

\begin{figure*}
\begin{center}
\includegraphics[width=\textwidth]{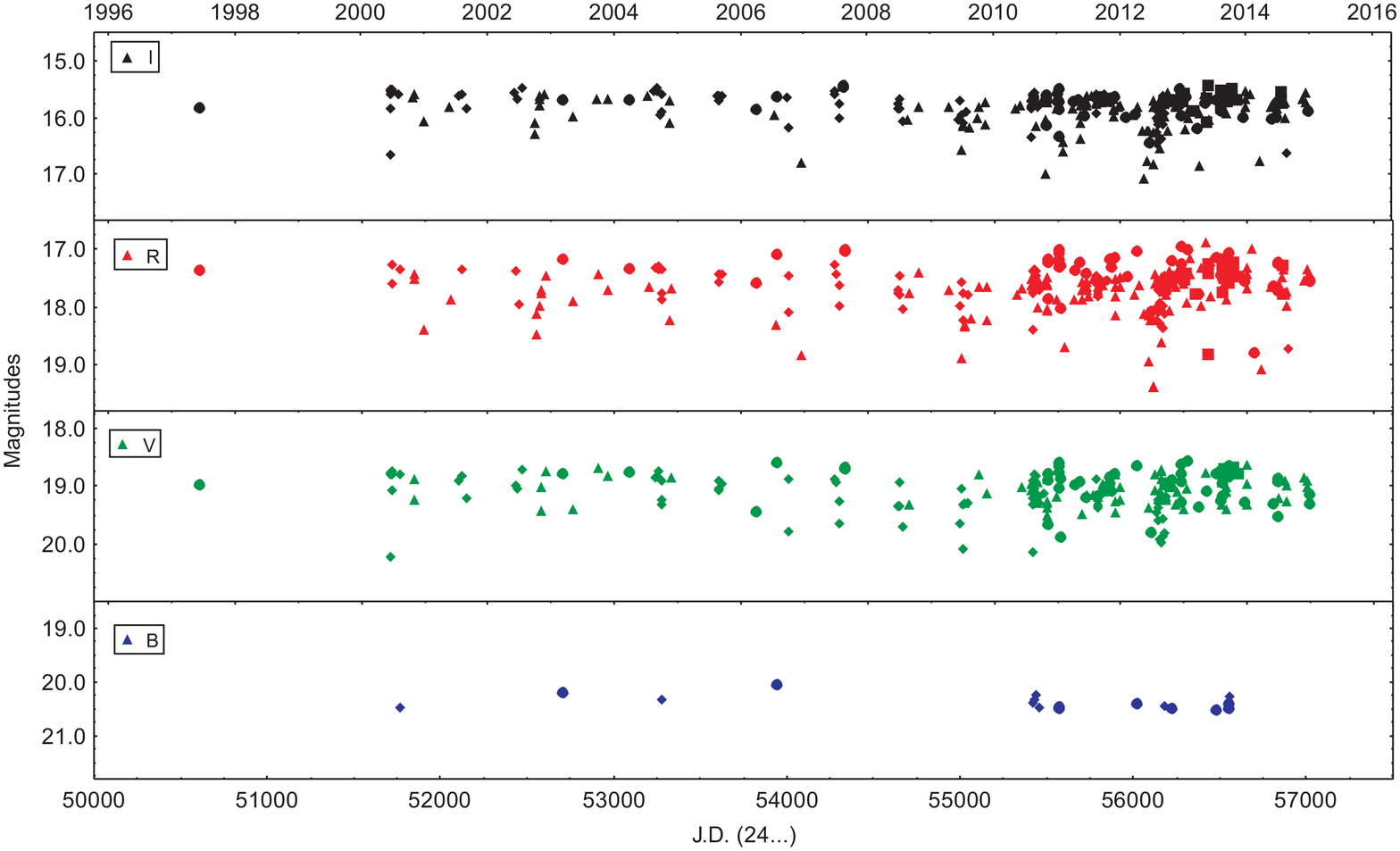}
\caption{CCD $IRVB$ light curves of FHO 28 for the period June 1997$-$December 2014}\label{Fig6}
\end{center}
\end{figure*}

The $BVRI$ light curves of FHO 28 from our CCD observations in the period 1997$-$2014 are shown in Fig.~\ref{Fig6}. In the figure, circles denote CCD photometric data taken with the 2-m RCC telescope, the other symbols used for the 1.3-m RC, the 50/70-cm Schmidt and the 60-cm Cassegrain telescopes are as in Fig.~\ref{Fig4}. The results of our long-term multicolor CCD observations of the star are summarized in Table~\ref{Tab4}. The columns have the same contents as in Table~\ref{Tab3}.

It can be seen from the Fig. 6 that FHO 28 spends most of the time at high light. The star exhibit strong irregular multiple fading in all bands with different amplitudes. 
The periods of decline in the brightness are relatively short, sometimes we have only one photometric point in the deep minimum.
The brightness of FHO 28 during the period of our observations 2000$-$2014 vary in the range 15.42$-$17.07 mag for $I$-band and 16.89$-$19.38 mag for $R$-band. The observed amplitudes are 1.65 mag for $I$- and 2.49 mag for $R$-bands, respectively. 
Because of the limit of our photometric data the deeper drops in the brightness of FHO 28 in $V$- and especially in $B$-bands are not registered. 
We have only some photometric points of the star in $B$-band collected with the 2-m RCC and the 1.3-m RC telescopes. Evidences of periodicity in the brightness variability of the star are not detected.

\begin{figure}
\begin{center}
\includegraphics[width=4cm]{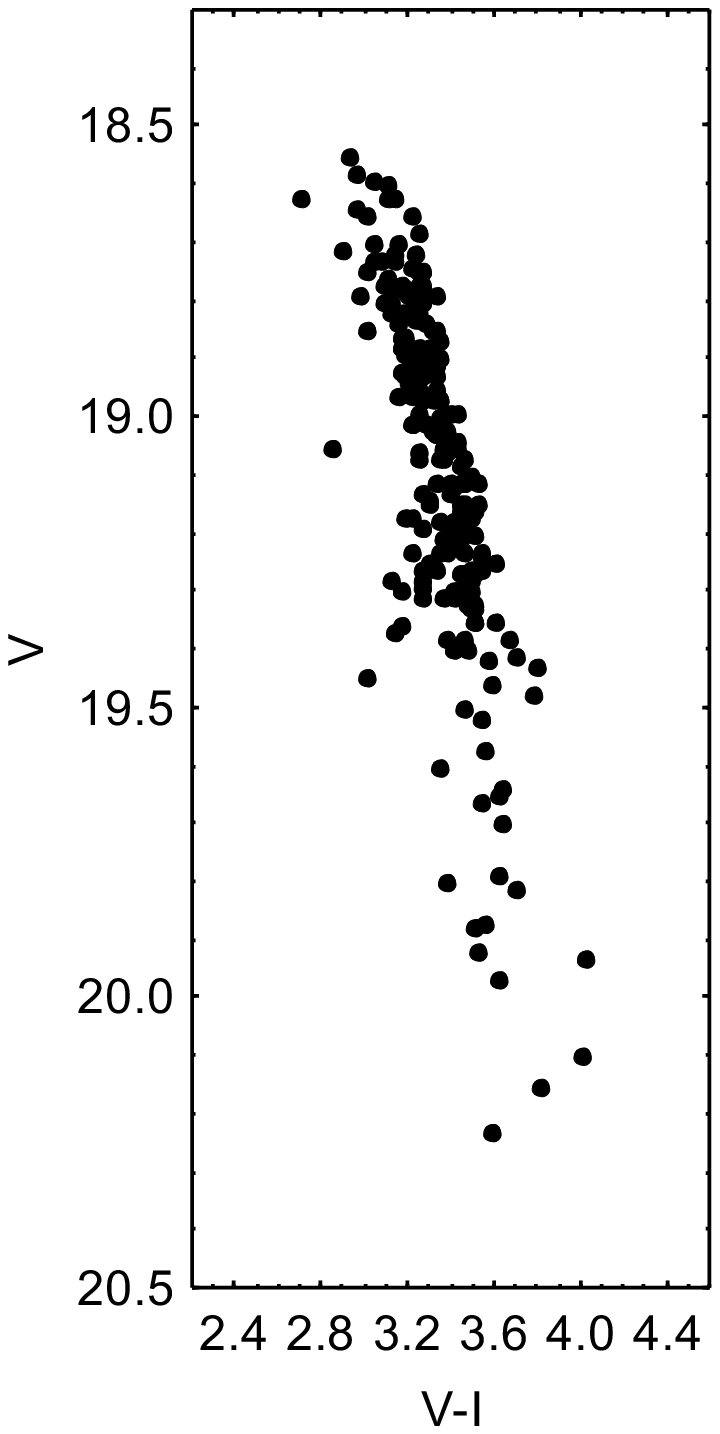}
\includegraphics[width=4cm]{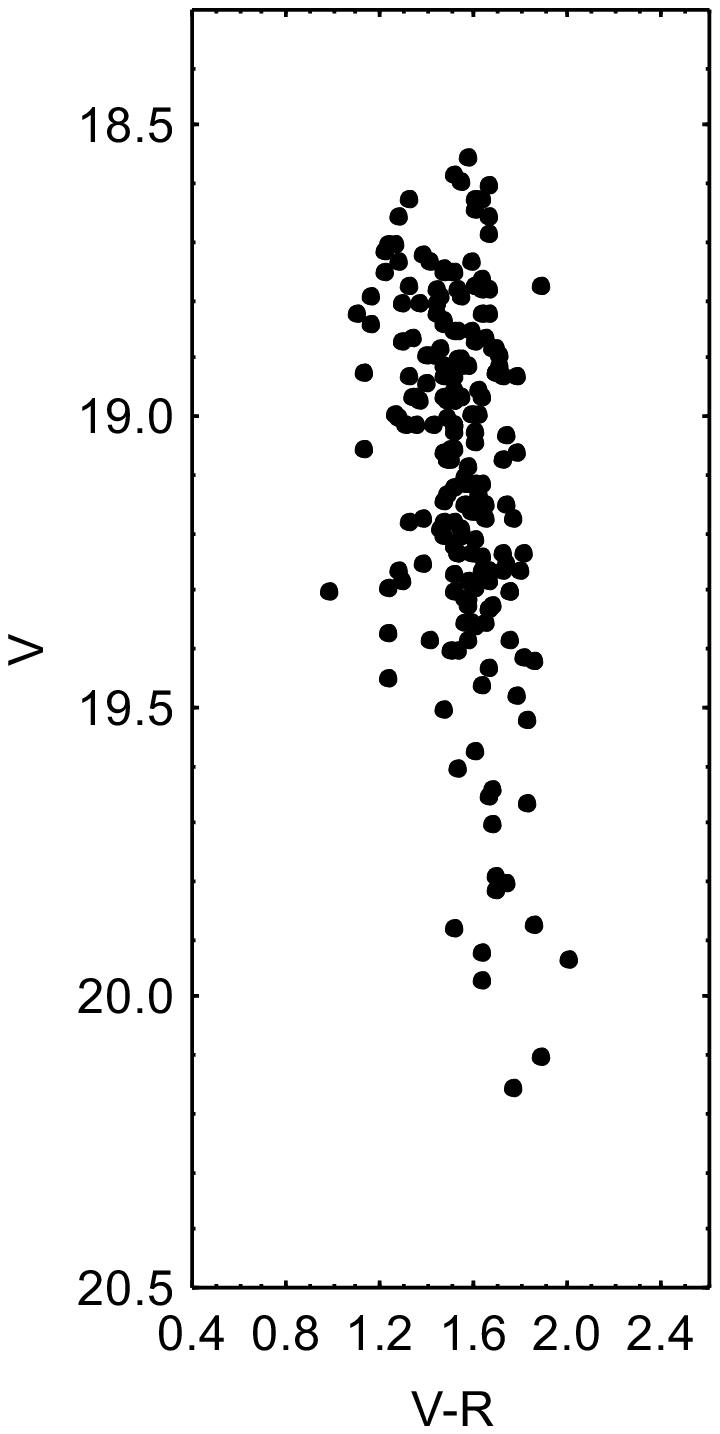}
\caption{Relationship between $V$ magnitude and the $V - I$ and $V - R$ color indexes of FHO 28 in the period June 1997$-$December 2014}\label{Fig7}
\end{center}
\end{figure}

Variability with largest amplitude of FHO 28 is observed in July 2012 (2.49 mag in $R$-band). 
The measured color indexes $V - I$ and $V - R$ versus the stellar $V$ magnitude during the period of our photometric observations are plotted in Fig.~\ref{Fig7}. 
Irrespective of the large amplitude of variability, the substantial bluing effect is not seen on the figure.
This means that the objects causing eclipses have a size significantly larger than the dust particles in the circumstellar environment of UXor variables.
It can be supposed that in this case we observe eclipses from a combination of multiple bodies rotating in orbits around the star.

\subsection{FHO 29}

The star FHO 29 (2MASS J20583974+4401328) is included in the list of YSO candidates by Guieu et~al. \shortcite{gui}. 
The authors measured $B$ = 18.288, $V$ = 16.892 and $I$ = 14.890 magnitudes of the star. 
Rebull et~al. \shortcite{rlm} classify FHO 29 as a flat spectrum source. 
Findeisen et~al. \shortcite{fin} presented $R$-band light curve of the star for the period 2011$-$2012 and described bursting events of the star.

\begin{figure*}
\begin{center}
\includegraphics[width=\textwidth]{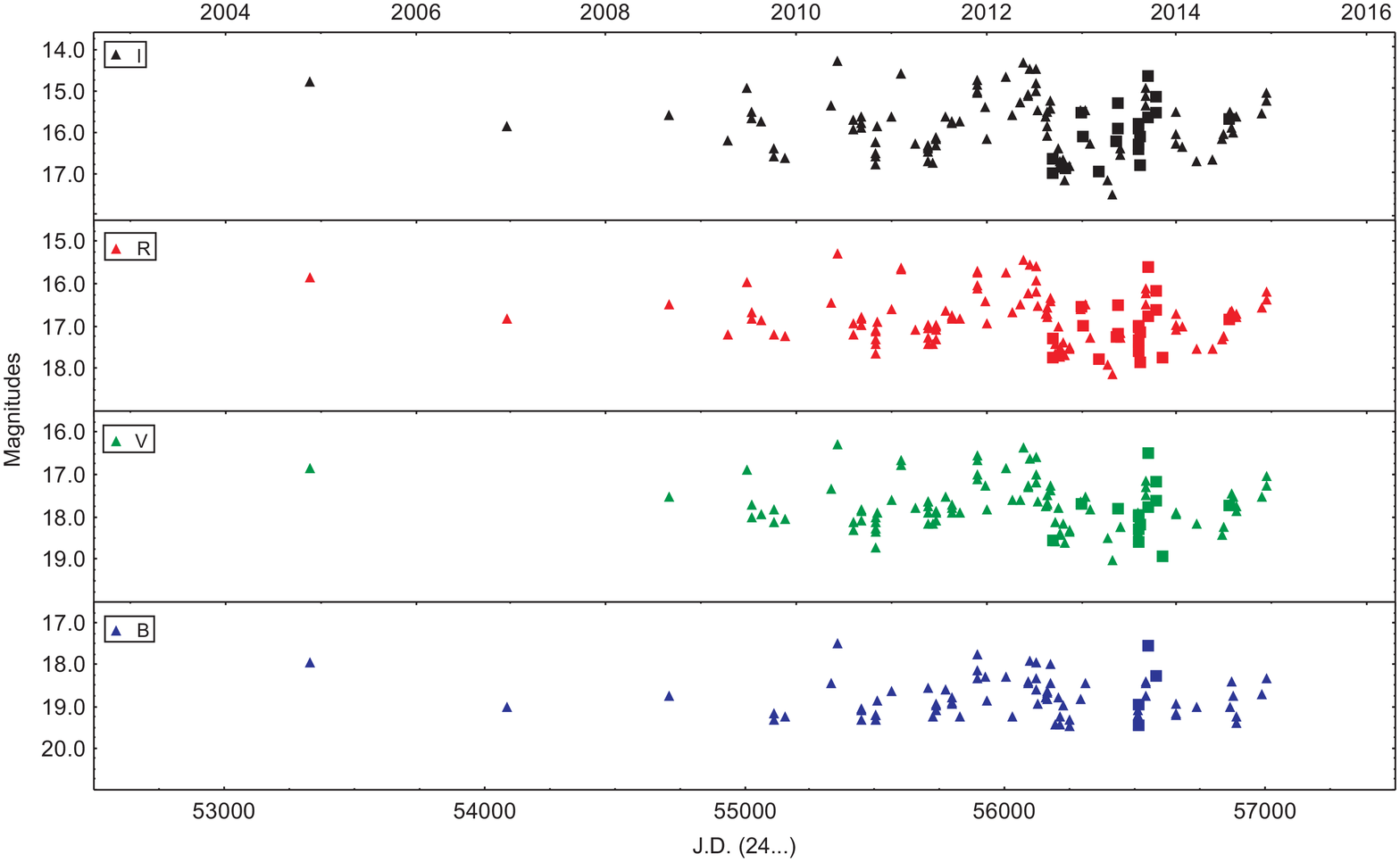}
\caption{CCD $IRVB$ light curves of FHO 29 for the period November 2004$-$December 2014}\label{Fig8}
\end{center}
\end{figure*}

The $BVRI$ light curves of FHO 29 from our CCD observations in the period 2004$-$2014 are shown in Fig.~\ref{Fig8}. The symbols used for the different telescopes are as in Fig.~\ref{Fig2}. The results of our long-term multicolor CCD observations of the star are summarized in Table~\ref{Tab5}. The columns have the same contents as in Table~\ref{Tab3}.

The brightness of FHO 29 during the period of our observations vary in the range 14.29$-$17.47 mag for $I$-band, 15.28$-$18.12 mag for $R$-band and 16.27$-$19.01 mag for $V$-band. The observed amplitudes are 3.18 mag for $I$-band, 2.84 mag for $R$-band and 2.74 mag for $V$-band in the same period, because the limit of our photometric data the deep fading events of FHO 29 in $B$-bands are not registered.

\begin{figure}
\begin{center}
\includegraphics[width=\columnwidth]{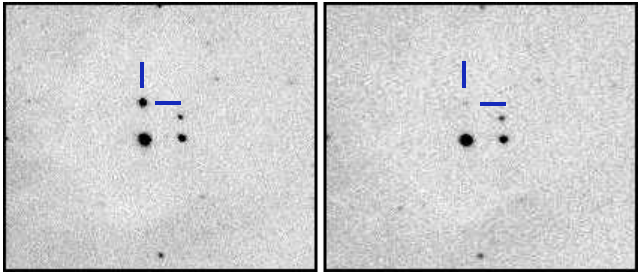}
\caption{The star FHO 29 at maximum (June 10, 2010, the left image) and at minimum light (May 2, 2013, the right image). The frames were taken in \textit{R}-band with Schmidt telescope of Rozhen NAO.}
\label{Fig9}
\end{center}
\end{figure}

It can be seen from Fig.~\ref{Fig8}, that until the end of 2011 FHO 29 exhibits irregular variations, then in the beginning of 2012 it began to increase its brightness and in mid-2012 the brightness of the star began sharply to drop, reaching amplitudes 3.10 mag for $I$-band, 2.70 mag for $R$-band and 2.60 mag for $V$-band in May 2012.
Several sharp increases in brightness for periods of several days or weeks were observed in 2013.
At the end of 2013 a gradual drop in brightness started following by a gradual rise by the end of 2014.
We can confirm that the long-term light curve of FHO 29 is dominated by burst events. 
Evidences of periodicity in the brightness variability of the star are not registered.
$R$-band images of FHO 29 at maximum (June 10, 2010) and at minimum light (May 02, 2013) are shown on Fig.~\ref{Fig9}. 

\begin{figure*}
\begin{center}
\includegraphics[width=4cm]{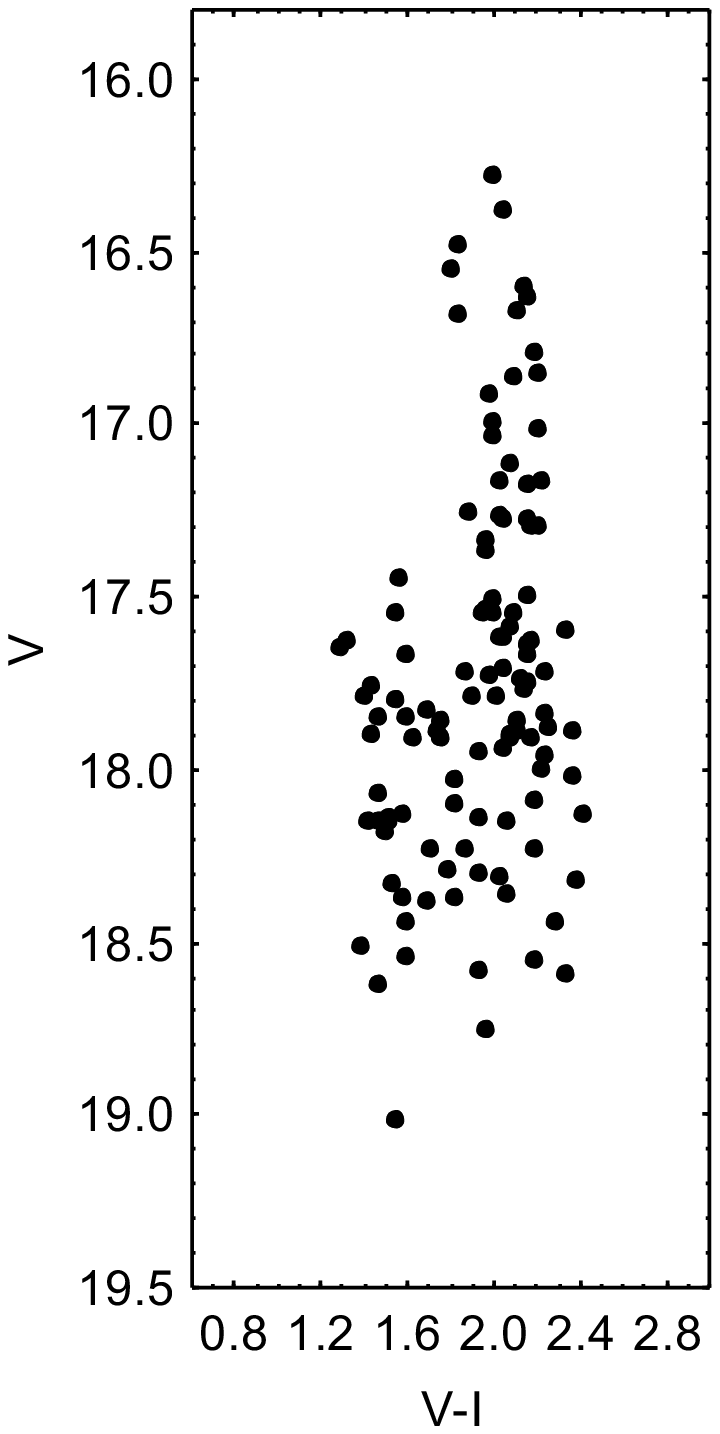}
\includegraphics[width=4cm]{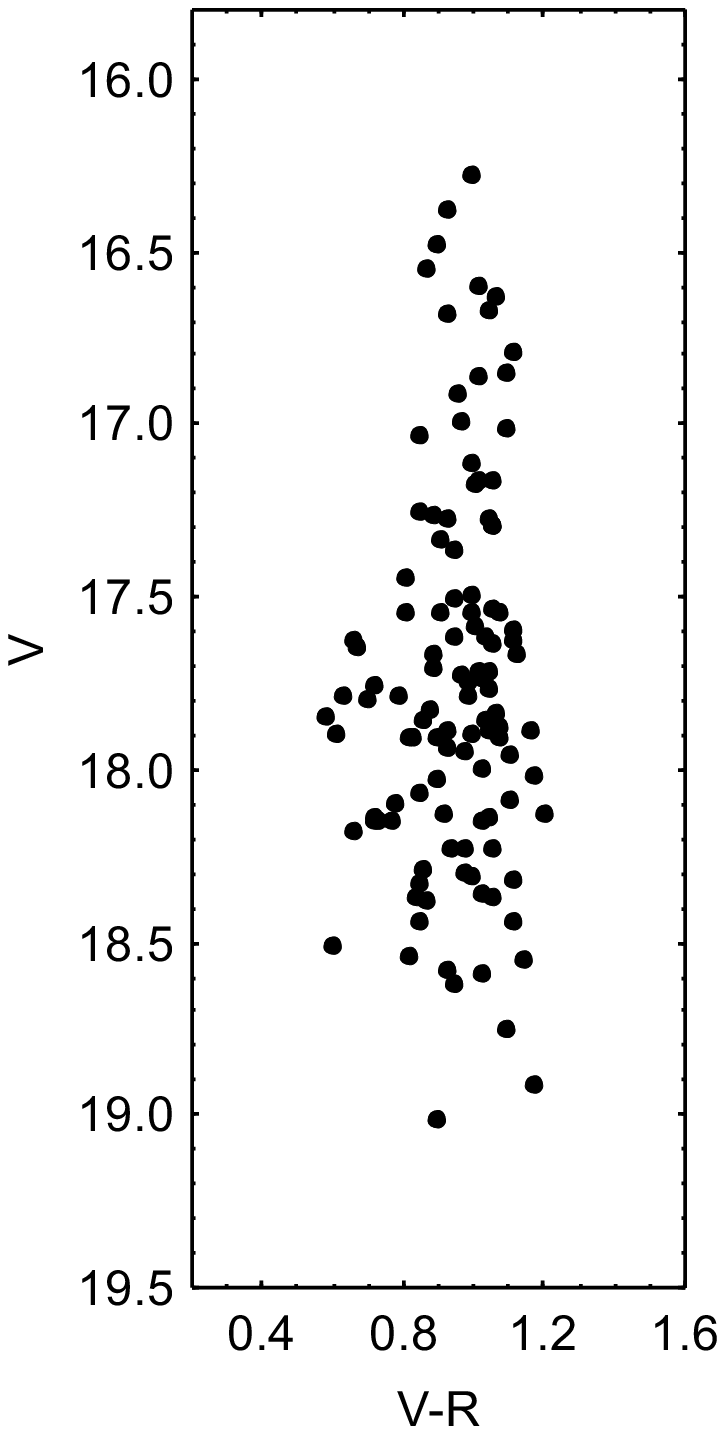}
\includegraphics[width=4cm]{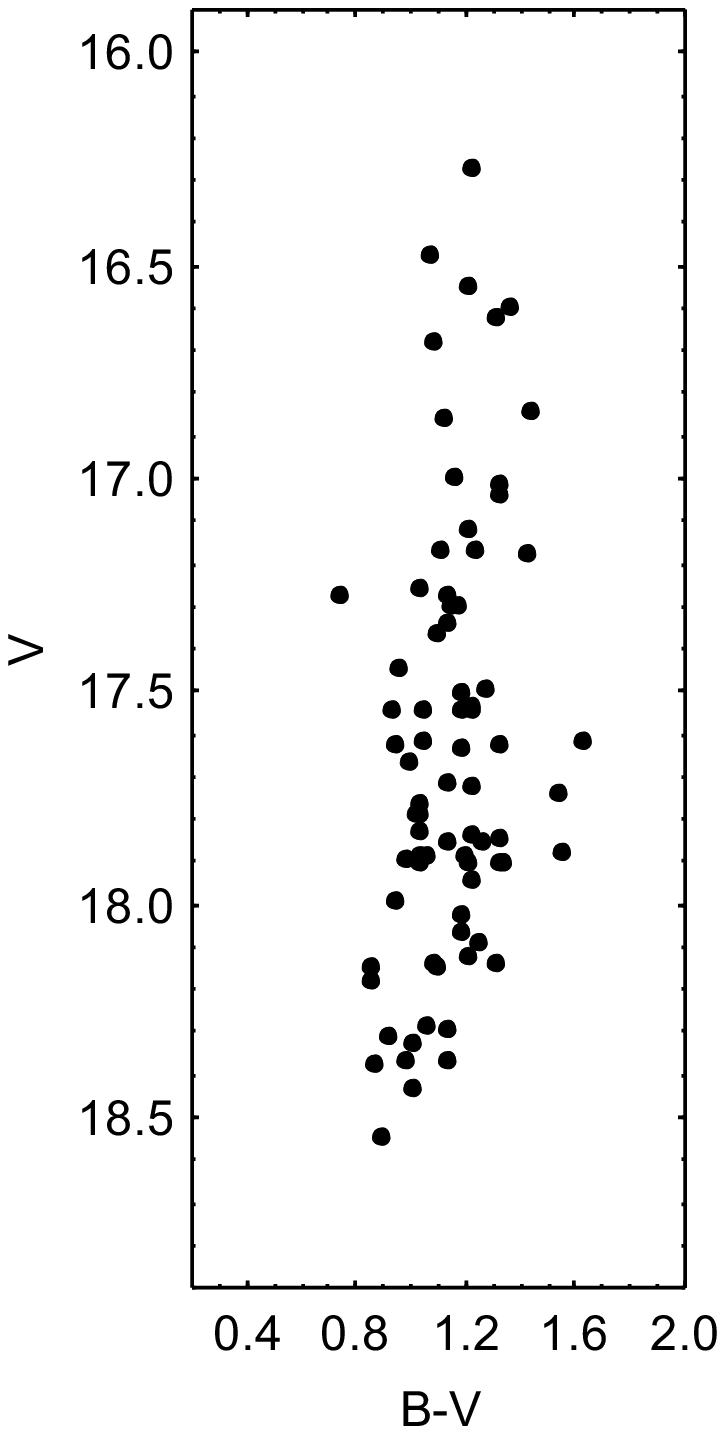}
\caption{Relationship between $V$ magnitude and the $V - I$, $V - R$ and $B - V$ color indexes of FHO 29 in the period November 2004$-$December 2014}\label{Fig10}
\end{center}
\end{figure*}

The measured color indexes $V - I$, $V - R$ and $B - V$ versus the stellar $V$ magnitude during the period of our observations are plotted in Fig.~\ref{Fig10}. 
In contrast to other objects from our study, for FHO 29 we observe an increase of the color indexes against an increase in the brightness of the star.
This relationship can be seen in Fig.~\ref{Fig8} also, the amplitude of the brightness in $I$-band is larger than the amplitudes in $R$- and $V$-bands, respectively.
Such a reverse of the color indexes is in a conflict with the bursting nature of the star.
On the other hand, the star can not be classified as UXor, as well as it spends more time at a low light.

\subsection{V1929 Cyg}

The star V1929 Cygni was discovered and classified as a flare star by Rosino et~al. \shortcite{ros}. 
The authors reported a flare event on September 4, 1972 when the brightness of the star increased from 16.50 mag $pg$ to 14.50 mag $pg$. 
The magnitudes and colors of V1929 Cyg in quiescence measured by Tsvetkov et~al. \shortcite{tst} are $V$ = 15.24 mag, $B-V$ = 0.92 mag, $V-R$ = 0.64 mag and $V-I$ = 1.28 mag. The authors provide spectrum of V1929 Cyg and estimate its spectral type as dK2-dK5, with H$\alpha$ in emission. 

Laugalys et~al. \shortcite{lys} measured $V$ = 15.111 mag and $V-U$ = 3.337 mag for V1929 Cyg and determined its photometric spectral type as ''K0V,e?''.
From a spectrum taken on October 22, 2007 by Corbally et~al. \shortcite{cor}, the authors identify faint H$\alpha$ emission line and OI, CaII and P9 lines. 
The spectral type is determined as G8e, but on the $J-H$/$H-K_{s}$ diagram the star lies close to the K dwarf sequence. 
The authors suggest that the energy distribution of the star, constructed from $Spitzer$ observations does not show any infrared emission and the authors conclude that the star seems to be a slightly reddened G dwarf with chromospheric activity.
The distance to V1929 Cyg determined by Corbally et~al. \shortcite{cor} is 580 pc, i.e., close to the distance of the NGC 7000/IC 5070 complex.

\begin{figure*}
\begin{center}
\includegraphics[width=\textwidth]{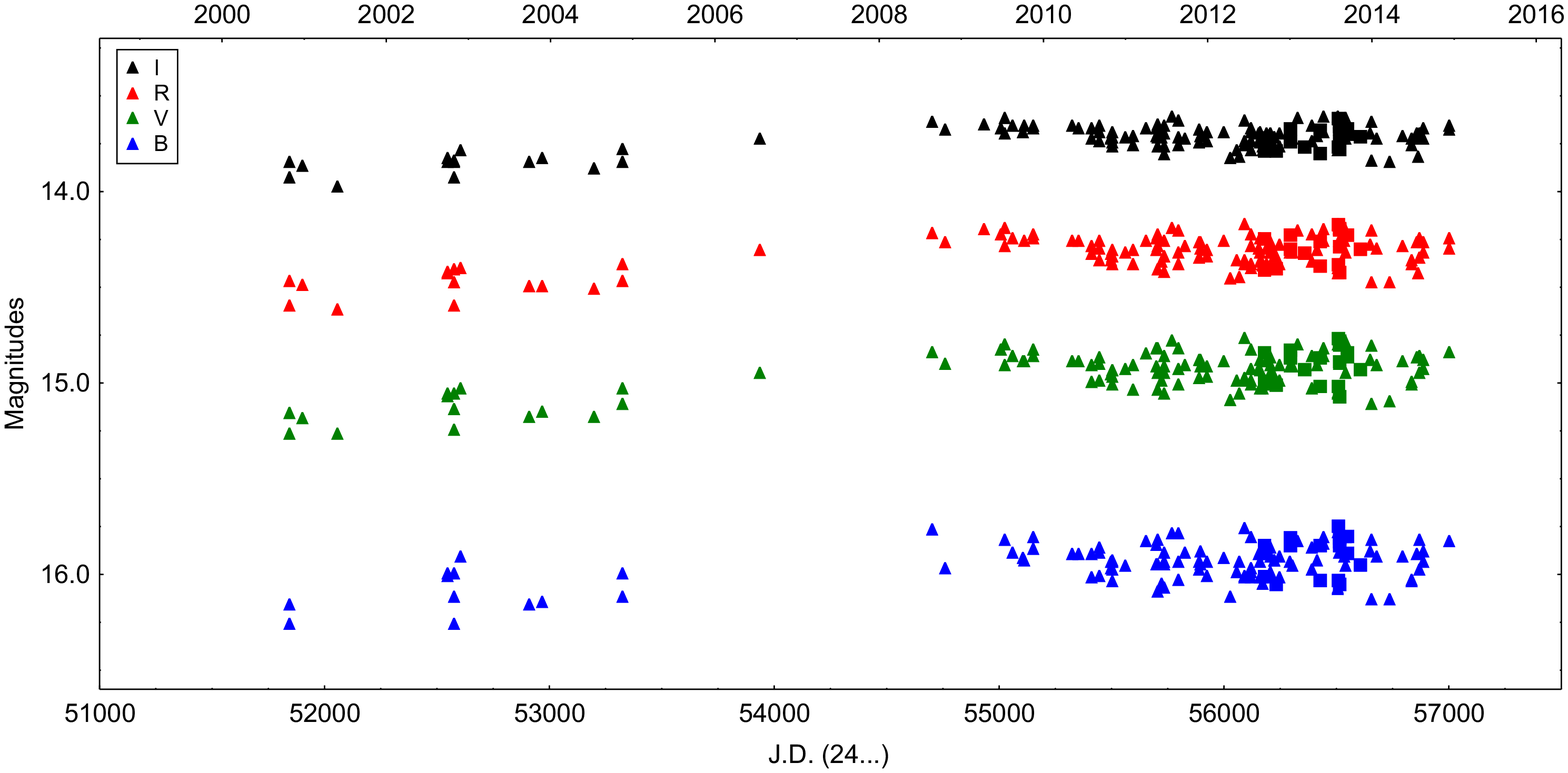}
\caption{CCD $IRVB$ light curves of V1929 for the period October 2000$-$December 2014}\label{Fig11}
\end{center}
\end{figure*}

The multicolor $BVRI$ light curves of V1929 Cyg from our CCD observations are shown in Fig.~\ref{Fig11}. The symbols used for the different telescopes are as in Fig.~\ref{Fig6}. The results of our long-term photometric observations are summarized in Table~\ref{Tab6}. 
The columns have the same contents as in Table~\ref{Tab3}.

From the Fig.~\ref{Fig11} can be seen that during the period of our observations V1929 Cyg shows irregular variability in all bands, but flare events are not registered.
The brightness of the star during the period 2000$-$2014 vary in the range of 13.61$-$13.98 mag for $I$-band, 14.17$-$14.62 mag for $R$-band, 14.77$-$15.27 mag for $V$-band and 15.75$-$16.26 mag for $B$-band. The observed amplitudes are $\Delta I$ = 0.37 mag, $\Delta R$ = 0.45 mag, $\Delta V$ = 0.50 mag and $\Delta B$ = 0.51 mag in the same period. Usually, such low amplitude variability is typical for low-mass WTTS.

\begin{figure*}
\begin{center}
\includegraphics[width=4cm]{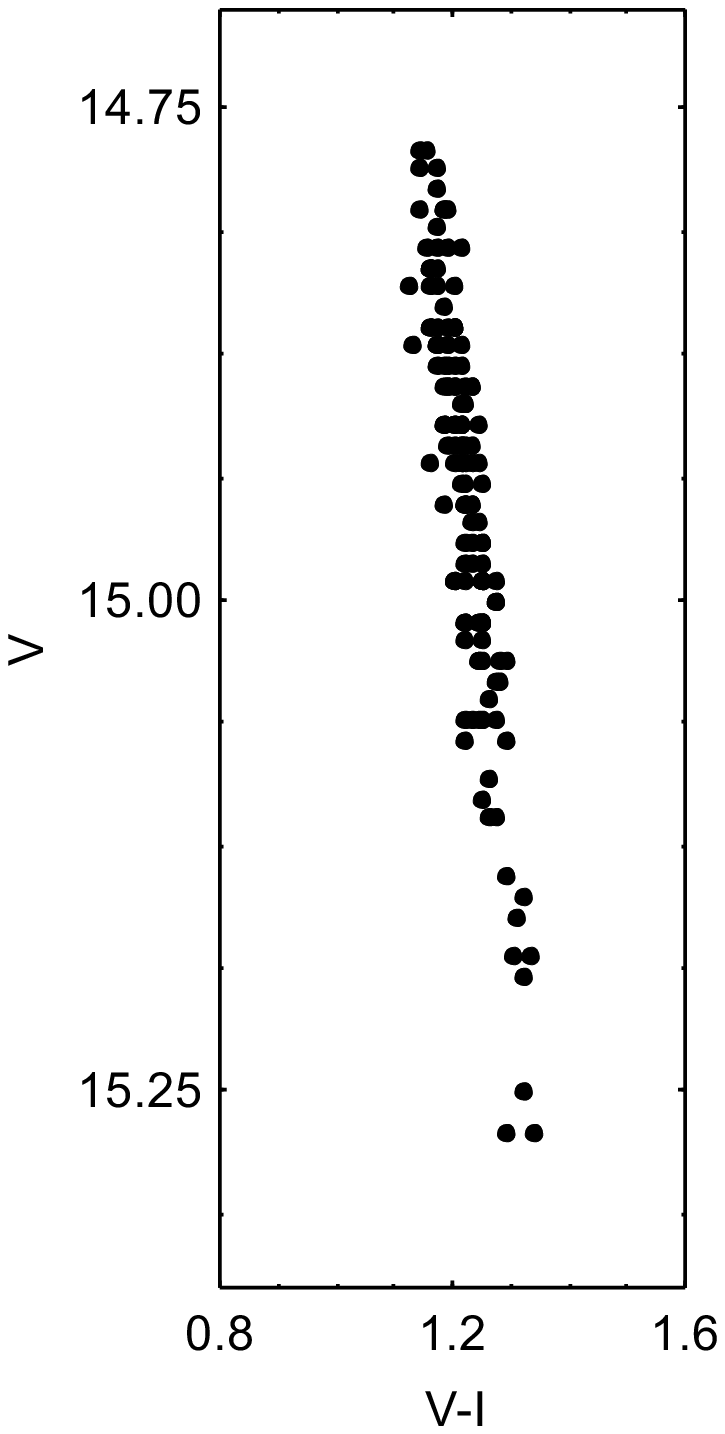}
\includegraphics[width=4cm]{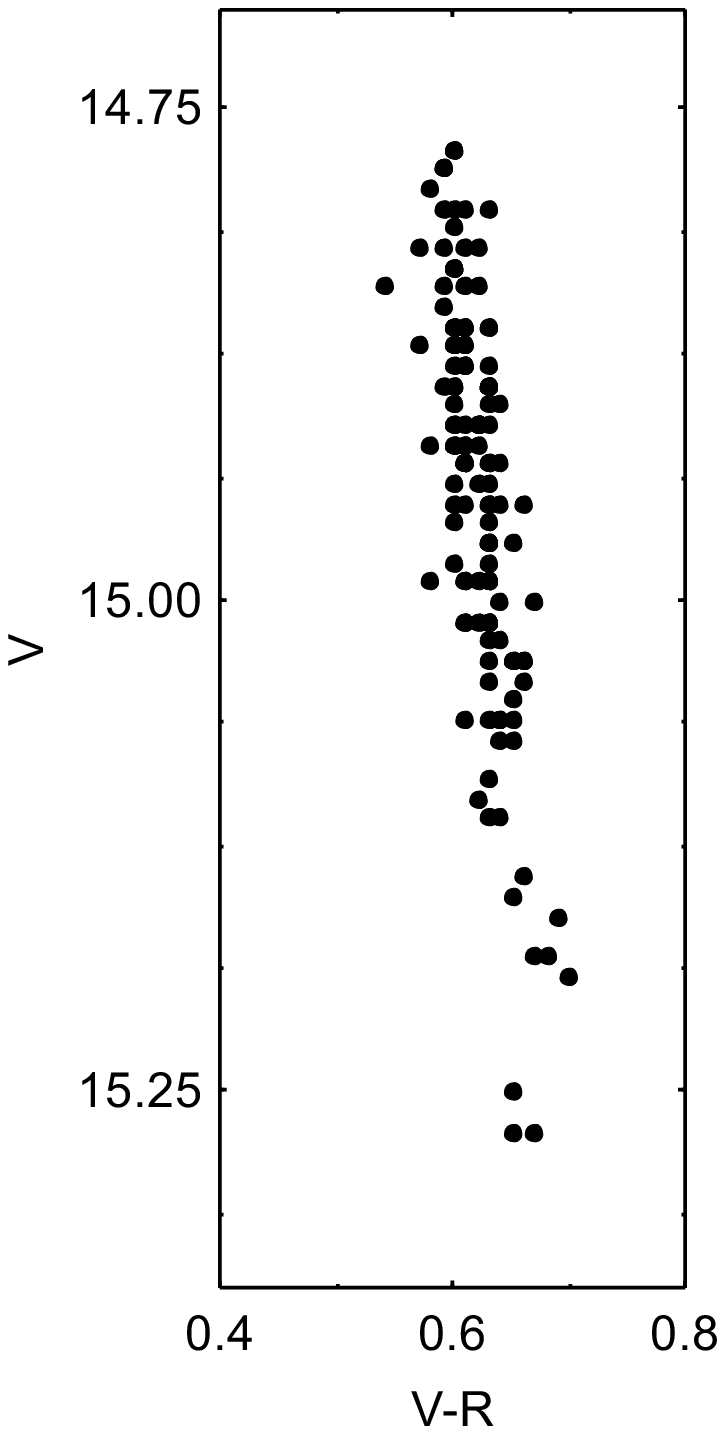}
\includegraphics[width=4cm]{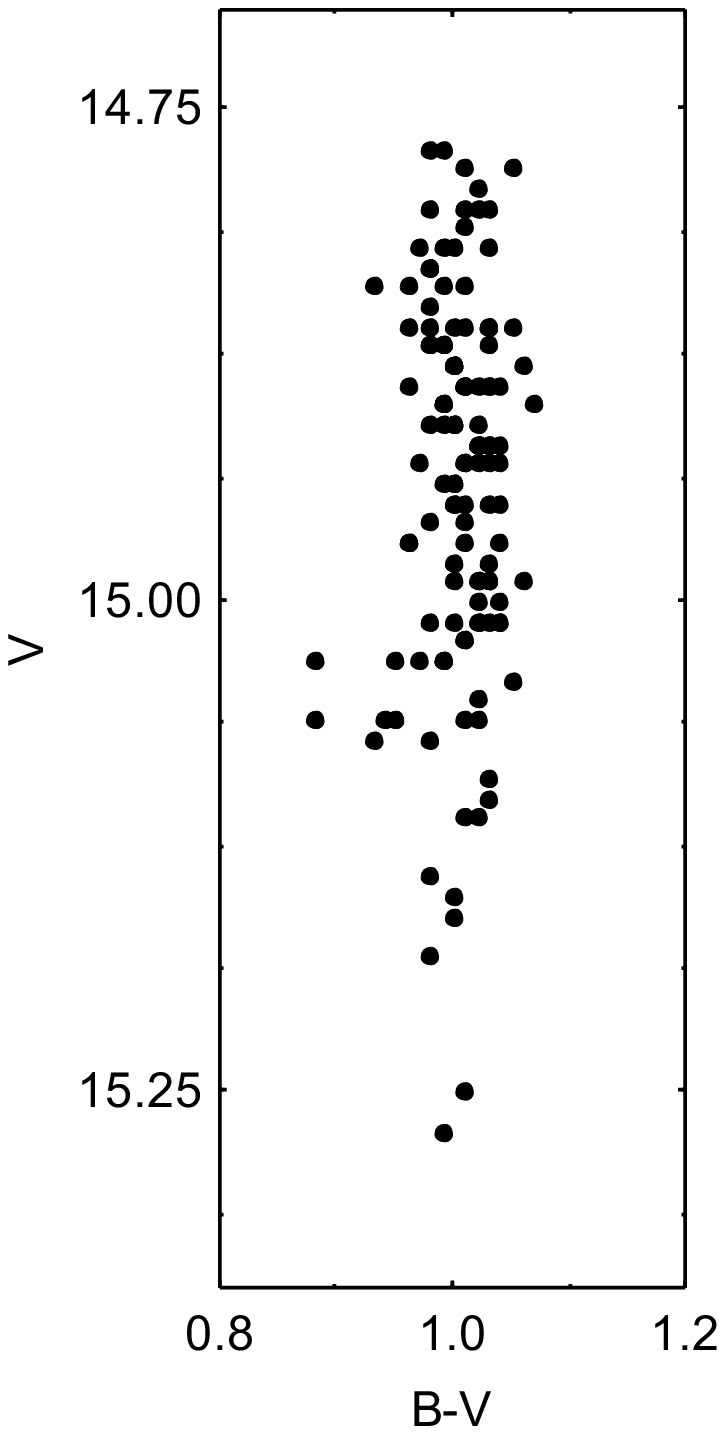}
\caption{Relationship between $V$ magnitude and the $V - I$, $V - R$ and $B - V$ color indexes of V1929 Cyg in the period October 2000$-$December 2014}\label{Fig12}
\end{center}
\end{figure*}

The measured color indexes $V - I$, $V - R$ and $B - V$ versus stellar $V$ magnitude during the period of our CCD observations are plotted in Fig.~\ref{Fig12}.
The color variations are typical of WTTS, stars with cool spots, whose variability is produced by rotation of the spotted surface.

We carried out a periodicity search using our data from Apr. 2012 to Dec. 2014 by \textit{PerSea} Version 2.6 (written by G. Maciejewski on the $ANOVA$ technique, Schwarzenberg-Cherny \shortcite{sch}) and \textit{Period04} \cite{len} softwares. Our time-series analysis of the data indicates a 0.426257 $\pm$ 0.000306 day period and led to the ephemeris: 
\begin{equation}\label{eq1}
JD (Max) = 2455113.949113 + 0.426257 * E.
\end{equation}

False Alarm Probability estimation was done by randomly deleting about 10\%$-$15\% of the data for about 30 times and then retuning the period of determination. The period and starting age ($T_{0}$) determination remain stable, even with a subsample with 30\% of the data removed.

\begin{figure}
\begin{center}
\includegraphics[width=\columnwidth]{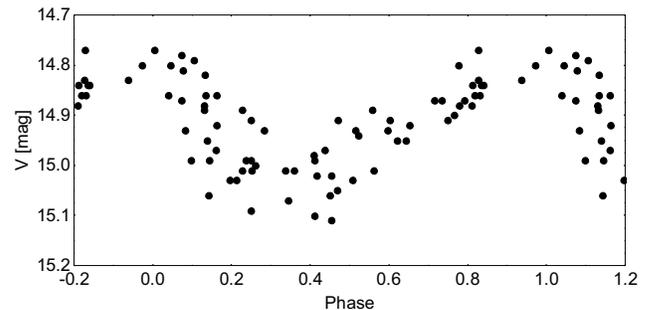}
\caption{$V$-band folded curve of V1929 Cyg}\label{Fig13}
\end{center}
\end{figure}

Figure~\ref{Fig13} exhibits the $V$-band folded light curve of V1929 Cyg according to the ephemeris (\ref{eq1}). 
The data obtained in $BRI$-bands show the same shape on periodicity. 
The found period is stable during time interval of several years and it is typical rotational period of young low-mass star.
The periodicity could be caused by rotation modulation of the dark spots on the stellar surface.
Such rotation period is extremely short for TTS, but still compatible with the shortest rotation periods of others WTTS \cite{hem, olz}.

\section{CONCLUSION}
We investigated the long-term photometric behavior of five pre-main sequence stars located in the region of ''Gulf of Mexico''. 
The stars FHO 26 and FHO 29 are likely CTTS showing large amplitude bursting events, while FHO 27 and FHO 28 have indications for variable extinction from the circumstellar environment. 
On the basis of the observed different shapes of the drops in the brightness we assume that there is likely to exist a variety of obscuration reasons. 
V1929 Cyg seems to be WTTS with extremely short rotation period and flare events from UV Cet type.

We are continuing to collect photometric observations of the field of ''Gulf of Mexico''. The long-term multicolor photometric observations of young stellar objects are very important for their exact classification. Further regular observations of the young stellar objects in the same field will be of a great importance and are encouraging.

\begin{acknowledgements}
The authors thank the Director of Skinakas Observatory Prof. I. Papamastorakis and Prof. I. Papadakis for the award of telescope time. The research has made use of the NASA's Astrophysics Data System Abstract Service and the SIMBAD database, operated at CDS, Strasbourg, France. This study was partly supported by the European Social Fund and the Ministry of Education and Science of Bulgaria under the contract BG051PO001-3.3.06-0047.
\end{acknowledgements}

\begin{appendix}

\section{PHOTOMETRIC CCD OBSERVATIONS AND DATA OF THE STARS FROM OUR STUDY}

\onecolumn
{\small
\begin{longtable}{ccccccc}
\caption{Photometric CCD observations and data of FHO 26 during the period October 2000$-$December 2014}\\
\hline\hline
\noalign{\smallskip}  
Date & J.D. (24...) & I & R & V & Tel & CCD\\
\noalign{\smallskip}  
\hline
\endfirsthead
\caption{Continued.}\\
\hline\hline
\noalign{\smallskip}  
Date & J.D. (24...) & I & R & V & Tel & CCD\\
\noalign{\smallskip}  
\hline
\noalign{\smallskip}  
\endhead
\hline
\label{Tab2}
\endfoot
\noalign{\smallskip}
29.10.2000	&	51847.241	&	15.97	&	18.05	&	19.15	&	Sch	&	ST-8	\\
30.10.2000	&	51848.300	&	16.02	&	17.99	&	19.21	&	Sch	&	ST-8	\\
24.12.2000	&	51903.243	&	15.99	&	17.89	&	19.16	&	Sch	&	ST-8	\\
27.05.2001	&	52057.425	&	15.96	&	17.93	&	19.14	&	Sch	&	ST-8	\\
03.10.2002	&	52551.364	&	16.04	&	18.05	&	19.10	&	Sch	&	ST-8	\\
04.10.2002	&	52552.394	&	16.16	&	18.32	&	-	&	Sch	&	ST-8	\\
29.10.2002	&	52577.288	&	15.96	&	18.00	&	19.30	&	Sch	&	ST-8	\\
30.10.2002	&	52578.284	&	16.01	&	18.01	&	19.36	&	Sch	&	ST-8	\\
01.11.2002	&	52580.193	&	16.16	&	18.28	&	-	&	Sch	&	ST-8	\\
06.03.2003	&	52765.517	&	16.03	&	18.05	&	19.34	&	Sch	&	ST-8	\\
27.09.2003	&	52910.375	&	16.00	&	18.04	&	19.50	&	Sch	&	ST-8	\\
25.11.2003	&	52969.192	&	16.00	&	18.06	&	19.28	&	Sch	&	ST-8	\\
17.07.2004	&	53203.607	&	15.94	&	17.87	&	18.99	&	Sch	&	ST-8	\\
19.07.2006	&	53936.474	&	16.05	&	18.54	&	-	&	Sch	&	ST-8	\\
23.10.2008	&	54763.202	&	16.04	&	17.99	&	-	&	Sch	&	STL-11	\\
28.06.2009	&	55011.458	&	16.12	&	18.00	&	-	&	Sch	&	FLI	\\
14.07.2009	&	55027.412	&	16.18	&	18.12	&	-	&	Sch	&	FLI	\\
15.07.2009	&	55028.380	&	16.09	&	-	&	-	&	Sch	&	FLI	\\
21.08.2009	&	55065.292	&	16.14	&	18.28	&	-	&	Sch	&	FLI	\\
06.10.2009	&	55111.330	&	16.05	&	-	&	18.84	&	Sch	&	FLI	\\
08.10.2009	&	55113.243	&	15.95	&	17.82	&	19.01	&	Sch	&	FLI	\\
20.11.2009	&	55156.197	&	16.01	&	17.90	&	19.23	&	Sch	&	FLI	\\
21.11.2009	&	55157.227	&	15.94	&	17.77	&	19.18	&	Sch	&	FLI	\\
13.05.2010	&	55330.436	&	15.99	&	17.90	&	19.46	&	Sch	&	FLI	\\
10.06.2010	&	55358.424	&	15.96	&	17.86	&	19.10	&	Sch	&	FLI	\\
06.08.2010	&	55415.466	&	15.99	&	17.67	&	18.75	&	Sch	&	FLI	\\
07.08.2010	&	55416.433	&	15.84	&	17.43	&	18.47	&	Sch	&	FLI	\\
07.09.2010	&	55447.420	&	15.97	&	17.75	&	19.00	&	Sch	&	FLI	\\
09.09.2010	&	55449.407	&	15.89	&	17.67	&	18.85	&	Sch	&	FLI	\\
31.10.2010	&	55501.195	&	15.84	&	17.53	&	18.62	&	Sch	&	FLI	\\
02.11.2010	&	55503.223	&	15.95	&	17.78	&	18.93	&	Sch	&	FLI	\\
03.11.2010	&	55504.195	&	16.00	&	17.84	&	18.81	&	Sch	&	FLI	\\
04.11.2010	&	55505.202	&	15.90	&	17.54	&	18.60	&	Sch	&	FLI	\\
05.11.2010	&	55506.226	&	15.92	&	17.58	&	18.86	&	Sch	&	FLI	\\
06.11.2010	&	55507.224	&	15.98	&	17.81	&	19.11	&	Sch	&	FLI	\\
01.01.2011	&	55563.169	&	15.92	&	17.75	&	19.02	&	Sch	&	FLI	\\
08.02.2011	&	55600.671	&	15.95	&	17.87	&	-	&	Sch	&	FLI	\\
04.04.2011	&	55656.509	&	15.89	&	17.64	&	-	&	Sch	&	FLI	\\
21.05.2011	&	55703.384	&	16.05	&	17.91	&	-	&	Sch	&	FLI	\\
22.05.2011	&	55704.430	&	16.03	&	17.75	&	-	&	Sch	&	FLI	\\
23.05.2011	&	55705.408	&	15.81	&	17.78	&	19.02	&	Sch	&	FLI	\\
24.05.2011	&	55706.382	&	15.99	&	17.92	&	-	&	Sch	&	FLI	\\
25.05.2011	&	55707.390	&	15.96	&	-	&	-	&	Sch	&	FLI	\\
09.06.2011	&	55722.425	&	15.92	&	17.56	&	-	&	Sch	&	FLI	\\
21.06.2011	&	55734.420	&	15.94	&	17.83	&	-	&	Sch	&	FLI	\\
22.06.2011	&	55735.452	&	16.02	&	17.86	&	18.81	&	Sch	&	FLI	\\
23.06.2011	&	55736.443	&	15.96	&	17.44	&	18.48	&	Sch	&	FLI	\\
24.06.2011	&	55737.443	&	15.97	&	17.66	&	18.72	&	Sch	&	FLI	\\
27.07.2011	&	55770.361	&	15.89	&	17.52	&	18.47	&	Sch	&	FLI	\\
23.08.2011	&	55797.323	&	16.02	&	17.88	&	19.30	&	Sch	&	FLI	\\
24.08.2011	&	55798.301	&	15.98	&	17.66	&	18.71	&	Sch	&	FLI	\\
25.08.2011	&	55799.318	&	15.87	&	17.70	&	18.57	&	Sch	&	FLI	\\
23.09.2011	&	55828.249	&	15.96	&	17.91	&	18.82	&	Sch	&	FLI	\\
27.11.2011	&	55893.176	&	15.91	&	17.84	&	-	&	Sch	&	FLI	\\
28.11.2011	&	55894.175	&	15.94	&	17.86	&	18.98	&	Sch	&	FLI	\\
29.11.2011	&	55895.167	&	15.88	&	17.91	&	19.06	&	Sch	&	FLI	\\
30.11.2011	&	55896.179	&	15.86	&	17.91	&	18.92	&	Sch	&	FLI	\\
29.12.2011	&	55925.169	&	15.98	&	17.95	&	18.89	&	Sch	&	FLI	\\
01.01.2012	&	55928.185	&	15.93	&	17.88	&	18.92	&	Sch	&	FLI	\\
16.03.2012	&	56003.563	&	15.88	&	17.70	&	-	&	Sch	&	FLI	\\
13.04.2012	&	56030.508	&	16.02	&	17.97	&	-	&	Sch	&	FLI	\\
12.05.2012	&	56060.415	&	15.94	&	-	&	-	&	Sch	&	FLI	\\
20.05.2012	&	56068.399	&	15.92	&	-	&	-	&	Sch	&	FLI	\\
12.06.2012	&	56091.398	&	16.02	&	17.96	&	19.06	&	Sch	&	FLI	\\
13.06.2012	&	56092.383	&	15.90	&	17.83	&	18.83	&	Sch	&	FLI	\\
17.06.2012	&	56096.398	&	15.98	&	18.01	&	-	&	Sch	&	FLI	\\
11.07.2012	&	56120.377	&	15.90	&	17.78	&	18.92	&	Sch	&	FLI	\\
12.07.2012	&	56121.338	&	15.90	&	17.85	&	18.83	&	Sch	&	FLI	\\
13.07.2012	&	56122.399	&	15.88	&	17.96	&	19.10	&	Sch	&	FLI	\\
14.07.2012	&	56123.387	&	16.06	&	18.10	&	-	&	Sch	&	FLI	\\
22.08.2012	&	56162.338	&	16.06	&	18.05	&	-	&	Sch	&	FLI	\\
04.09.2012	&	56175.426	&	15.90	&	17.75	&	18.79	&	Sch	&	FLI	\\
05.09.2012	&	56176.272	&	15.94	&	17.89	&	19.06	&	Sch	&	FLI	\\
08.09.2012	&	56179.272	&	16.02	&	18.03	&	-	&	60-cm	&	FLI	\\
09.09.2012	&	56180.364	&	15.97	&	17.84	&	-	&	60-cm	&	FLI	\\
23.09.2012	&	56194.322	&	16.04	&	17.95	&	19.09	&	Sch	&	FLI	\\
07.10.2012	&	56208.210	&	15.98	&	17.92	&	18.91	&	Sch	&	FLI	\\
08.10.2012	&	56209.204	&	15.94	&	17.71	&	-	&	Sch	&	FLI	\\
09.10.2012	&	56210.193	&	15.99	&	18.04	&	19.30	&	Sch	&	FLI	\\
10.10.2012	&	56211.461	&	15.81	&	17.65	&	-	&	Sch	&	FLI	\\
25.10.2012	&	56226.240	&	15.95	&	17.80	&	19.02	&	Sch	&	FLI	\\
26.10.2012	&	56227.401	&	15.95	&	17.90	&	-	&	Sch	&	FLI	\\
17.11.2012	&	56249.172	&	15.92	&	17.89	&	19.26	&	Sch	&	FLI	\\
18.11.2012	&	56250.181	&	16.06	&	18.05	&	-	&	Sch	&	FLI	\\
31.12.2012	&	56293.254	&	16.01	&	18.01	&	19.23	&	Sch	&	FLI	\\
16.01.2013	&	56309.243	&	15.92	&	17.82	&	-	&	Sch	&	FLI	\\
10.04.2013	&	56392.510	&	15.92	&	-	&	-	&	Sch	&	FLI	\\
11.04.2013	&	56394.482	&	16.10	&	18.28	&	-	&	Sch	&	FLI	\\
04.08.2013	&	56509.288	&	16.02	&	17.97	&	-	&	Sch	&	FLI	\\
05.08.2013	&	56510.386	&	16.11	&	18.20	&	19.32	&	Sch	&	FLI	\\
06.08.2013	&	56511.411	&	16.09	&	18.34	&	19.09	&	Sch	&	FLI	\\
07.08.2013	&	56512.398	&	16.02	&	18.10	&	19.10	&	Sch	&	FLI	\\
04.09.2013	&	56540.274	&	15.88	&	17.61	&	18.55	&	Sch	&	FLI	\\
28.12.2013	&	56655.200	&	16.01	&	18.04	&	-	&	Sch	&	FLI	\\
29.12.2013	&	56656.180	&	15.89	&	17.81	&	18.73	&	Sch	&	FLI	\\
23.01.2014	&	56681.189	&	15.98	&	17.74	&	18.68	&	Sch	&	FLI	\\
22.03.2014	&	56738.590	&	15.92	&	17.73	&	18.75	&	Sch	&	FLI	\\
21.05.2014	&	56799.425	&	15.93	&	17.79	&	-	&	Sch	&	FLI	\\
28.06.2014	&	56837.417	&	15.99	&	17.94	&	-	&	Sch	&	FLI	\\
29.06.2014	&	56838.398	&	15.89	&	17.84	&	19.18	&	Sch	&	FLI	\\
24.07.2014	&	56863.339	&	15.93	&	17.86	&	18.90	&	Sch	&	FLI	\\
25.07.2014	&	56864.356	&	16.01	&	-	&	-	&	Sch	&	FLI	\\
03.08.2014	&	56873.386	&	15.87	&	17.77	&	18.70	&	Sch	&	FLI	\\
04.08.2014	&	56874.407	&	15.87	&	17.80	&	18.92	&	Sch	&	FLI	\\
18.08.2014	&	56888.450	&	15.95	&	17.68	&	18.76	&	Sch	&	FLI	\\
19.08.2014	&	56889.365	&	15.83	&	17.54	&	18.61	&	Sch	&	FLI	\\
13.12.2014	&	57005.181	&	16.05	&	17.87	&	18.79	&	Sch	&	FLI	\\
14.12.2014	&	57006.220	&	15.87	&	17.76	&	18.57	&	Sch	&	FLI	\\
\hline \hline
\end{longtable}}

{\small
\begin{longtable}{cccccccc}
\caption{Photometric CCD observations and data of FHO 27 during the period October 2000$-$December 2014}\\
\hline\hline
\noalign{\smallskip}  
Date & J.D. (24...) & I & R & V & B & Tel & CCD\\
\noalign{\smallskip}  
\hline
\endfirsthead
\caption{Continued.}\\
\hline\hline
\noalign{\smallskip}  
Date & J.D. (24...) & I & R & V & B & Tel & CCD\\
\noalign{\smallskip}  
\hline
\noalign{\smallskip}  
\endhead
\hline
\label{Tab3}
\endfoot
\noalign{\smallskip}
29.10.2000	&	51847.241	&	14.70	&	16.05	&	17.21	&	18.55	&	Sch	&	ST-8	\\
30.10.2000	&	51848.300	&	14.63	&	15.99	&	17.16	&	18.60	&	Sch	&	ST-8	\\
24.12.2000	&	51903.243	&	14.54	&	15.88	&	17.02	&	-	&	Sch	&	ST-8	\\
27.05.2001	&	52057.425	&	14.64	&	16.00	&	17.24	&	-	&	Sch	&	ST-8	\\
03.10.2002	&	52551.364	&	14.79	&	16.13	&	17.43	&	19.19	&	Sch	&	ST-8	\\
04.10.2002	&	52552.394	&	14.78	&	16.16	&	17.44	&	18.64	&	Sch	&	ST-8	\\
29.10.2002	&	52577.288	&	14.36	&	15.58	&	16.68	&	17.84	&	Sch	&	ST-8	\\
30.10.2002	&	52578.284	&	14.42	&	15.73	&	16.84	&	18.29	&	Sch	&	ST-8	\\
01.11.2002	&	52580.193	&	14.69	&	16.07	&	17.26	&	19.23	&	Sch	&	ST-8	\\
28.11.2002	&	52607.212	&	14.70	&	16.03	&	17.18	&	18.53	&	Sch	&	ST-8	\\
06.03.2003	&	52765.517	&	15.08	&	16.55	&	17.78	&	19.42	&	Sch	&	ST-8	\\
27.09.2003	&	52910.375	&	14.97	&	16.37	&	17.69	&	19.10	&	Sch	&	ST-8	\\
25.11.2003	&	52969.192	&	14.89	&	16.30	&	17.55	&	18.86	&	Sch	&	ST-8	\\
17.07.2004	&	53203.607	&	14.79	&	16.14	&	17.30	&	-	&	Sch	&	ST-8	\\
18.11.2004	&	53328.231	&	14.54	&	15.84	&	17.02	&	18.40	&	Sch	&	ST-8	\\
20.11.2004	&	53330.273	&	14.68	&	16.04	&	17.15	&	18.91	&	Sch	&	ST-8	\\
19.07.2006	&	53936.474	&	14.96	&	16.39	&	17.73	&	-	&	Sch	&	ST-8	\\
16.12.2006	&	54086.204	&	14.95	&	16.17	&	-	&	-	&	Sch	&	ST-8	\\
23.07.2007	&	54305.319	&	14.74	&	16.01	&	17.25	&	18.80	&	1.3-m	&	ANDOR	\\
24.07.2007	&	54306.315	&	14.58	&	15.82	&	17.02	&	18.48	&	1.3-m	&	ANDOR	\\
28.08.2008	&	54707.312	&	14.69	&	15.81	&	16.94	&	18.46	&	Sch	&	STL-11	\\
23.10.2008	&	54763.202	&	14.56	&	15.82	&	17.05	&	-	&	Sch	&	STL-11	\\
16.04.2009	&	54938.551	&	14.92	&	16.22	&	17.42	&	19.19	&	Sch	&	STL-11	\\
26.06.2009	&	55009.521	&	14.82	&	16.12	&	17.37	&	18.95	&	1.3-m	&	ANDOR	\\
28.06.2009	&	55011.458	&	15.16	&	16.56	&	17.92	&	-	&	Sch	&	FLI	\\
04.07.2009	&	55016.500	&	14.85	&	16.16	&	17.40	&	18.98	&	1.3-m	&	ANDOR	\\
09.07.2009	&	55022.502	&	14.83	&	16.17	&	17.54	&	19.62	&	1.3-m	&	ANDOR	\\
14.07.2009	&	55027.412	&	14.85	&	16.16	&	17.37	&	-	&	Sch	&	FLI	\\
15.07.2009	&	55028.380	&	14.80	&	16.22	&	17.45	&	-	&	Sch	&	FLI	\\
21.08.2009	&	55065.292	&	15.27	&	16.74	&	18.22	&	-	&	Sch	&	FLI	\\
06.10.2009	&	55111.330	&	-	&	17.66	&	19.12	&	-	&	Sch	&	FLI	\\
08.10.2009	&	55113.243	&	15.13	&	16.52	&	17.85	&	19.30	&	Sch	&	FLI	\\
20.11.2009	&	55156.197	&	15.74	&	17.13	&	18.51	&	-	&	Sch	&	FLI	\\
21.11.2009	&	55157.227	&	15.66	&	16.99	&	18.35	&	-	&	Sch	&	FLI	\\
13.05.2010	&	55330.436	&	14.72	&	16.06	&	17.35	&	19.13	&	Sch	&	FLI	\\
10.06.2010	&	55358.424	&	14.80	&	16.15	&	17.48	&	-	&	Sch	&	FLI	\\
06.08.2010	&	55415.466	&	14.62	&	15.91	&	17.16	&	18.64	&	Sch	&	FLI	\\
07.08.2010	&	55416.433	&	14.63	&	15.95	&	17.24	&	18.84	&	Sch	&	FLI	\\
07.09.2010	&	55447.420	&	14.73	&	16.03	&	17.31	&	18.96	&	Sch	&	FLI	\\
08.09.2010	&	55448.326	&	14.65	&	15.99	&	17.23	&	18.88	&	Sch	&	FLI	\\
09.09.2010	&	55449.407	&	14.60	&	15.87	&	17.11	&	18.61	&	Sch	&	FLI	\\
31.10.2010	&	55501.195	&	14.51	&	15.84	&	17.06	&	18.54	&	Sch	&	FLI	\\
02.11.2010	&	55503.223	&	14.42	&	15.67	&	16.82	&	18.27	&	Sch	&	FLI	\\
03.11.2010	&	55504.195	&	14.50	&	15.79	&	16.96	&	18.59	&	Sch	&	FLI	\\
04.11.2010	&	55505.202	&	14.44	&	15.70	&	16.87	&	18.34	&	Sch	&	FLI	\\
05.11.2010	&	55506.226	&	14.59	&	15.89	&	17.11	&	18.61	&	Sch	&	FLI	\\
06.11.2010	&	55507.224	&	14.58	&	15.89	&	17.11	&	18.72	&	Sch	&	FLI	\\
01.01.2011	&	55563.169	&	14.85	&	16.18	&	17.45	&	18.89	&	Sch	&	FLI	\\
07.02.2011	&	55600.202	&	14.89	&	16.19	&	17.51	&	-	&	Sch	&	FLI	\\
08.02.2011	&	55600.671	&	14.82	&	16.18	&	17.50	&	-	&	Sch	&	FLI	\\
04.04.2011	&	55656.509	&	14.56	&	15.83	&	17.05	&	18.42	&	Sch	&	FLI	\\
21.05.2011	&	55703.384	&	14.72	&	16.01	&	17.20	&	18.78	&	Sch	&	FLI	\\
22.05.2011	&	55704.430	&	14.71	&	16.04	&	17.27	&	-	&	Sch	&	FLI	\\
23.05.2011	&	55705.408	&	14.77	&	16.10	&	17.24	&	19.08	&	Sch	&	FLI	\\
24.05.2011	&	55706.382	&	14.78	&	16.13	&	17.31	&	-	&	Sch	&	FLI	\\
25.05.2011	&	55707.390	&	14.77	&	16.02	&	17.26	&	-	&	Sch	&	FLI	\\
09.06.2011	&	55722.425	&	14.60	&	15.88	&	17.19	&	18.48	&	Sch	&	FLI	\\
21.06.2011	&	55734.420	&	14.71	&	15.98	&	17.24	&	18.52	&	Sch	&	FLI	\\
22.06.2011	&	55735.452	&	14.91	&	16.31	&	17.62	&	19.44	&	Sch	&	FLI	\\
23.06.2011	&	55736.443	&	14.92	&	16.32	&	17.66	&	-	&	Sch	&	FLI	\\
24.06.2011	&	55737.443	&	14.80	&	16.15	&	17.39	&	19.14	&	Sch	&	FLI	\\
27.07.2011	&	55770.361	&	14.73	&	16.07	&	17.32	&	18.92	&	Sch	&	FLI	\\
23.08.2011	&	55797.323	&	14.78	&	16.19	&	17.45	&	-	&	Sch	&	FLI	\\
24.08.2011	&	55798.301	&	14.63	&	15.95	&	17.23	&	18.76	&	Sch	&	FLI	\\
25.08.2011	&	55799.318	&	14.43	&	15.73	&	16.88	&	18.43	&	Sch	&	FLI	\\
23.09.2011	&	55828.249	&	14.78	&	16.14	&	17.49	&	19.16	&	Sch	&	FLI	\\
27.11.2011	&	55893.176	&	16.31	&	17.62	&	-	&	-	&	Sch	&	FLI	\\
28.11.2011	&	55894.175	&	16.05	&	17.52	&	18.78	&	-	&	Sch	&	FLI	\\
29.11.2011	&	55895.167	&	16.00	&	17.40	&	18.85	&	-	&	Sch	&	FLI	\\
30.11.2011	&	55896.179	&	16.19	&	17.65	&	19.07	&	-	&	Sch	&	FLI	\\
29.12.2011	&	55925.169	&	15.19	&	16.66	&	17.99	&	-	&	Sch	&	FLI	\\
01.01.2012	&	55928.185	&	15.28	&	16.72	&	18.12	&	-	&	Sch	&	FLI	\\
16.03.2012	&	56003.563	&	14.52	&	15.76	&	16.97	&	18.37	&	Sch	&	FLI	\\
13.04.2012	&	56030.508	&	15.60	&	17.09	&	18.80	&	-	&	Sch	&	FLI	\\
12.05.2012	&	56060.415	&	15.43	&	16.80	&	18.01	&	-	&	Sch	&	FLI	\\
20.05.2012	&	56068.399	&	15.37	&	16.71	&	-	&	-	&	Sch	&	FLI	\\
12.06.2012	&	56091.398	&	15.94	&	17.43	&	18.85	&	-	&	Sch	&	FLI	\\
13.06.2012	&	56092.383	&	15.70	&	17.17	&	18.55	&	-	&	Sch	&	FLI	\\
17.06.2012	&	56096.398	&	15.69	&	17.20	&	-	&	-	&	Sch	&	FLI	\\
11.07.2012	&	56120.377	&	15.70	&	17.22	&	18.59	&	-	&	Sch	&	FLI	\\
12.07.2012	&	56121.338	&	15.31	&	16.69	&	17.98	&	-	&	Sch	&	FLI	\\
13.07.2012	&	56122.399	&	15.49	&	16.92	&	18.20	&	-	&	Sch	&	FLI	\\
14.07.2012	&	56123.387	&	15.34	&	16.79	&	18.06	&	-	&	Sch	&	FLI	\\
19.08.2012	&	56159.351	&	15.69	&	17.21	&	18.61	&	-	&	Sch	&	FLI	\\
20.08.2012	&	56160.329	&	15.74	&	17.28	&	18.66	&	-	&	Sch	&	FLI	\\
21.08.2012	&	56161.346	&	15.65	&	17.12	&	18.44	&	-	&	Sch	&	FLI	\\
22.08.2012	&	56162.338	&	15.58	&	17.16	&	18.60	&	-	&	Sch	&	FLI	\\
04.09.2012	&	56175.426	&	15.64	&	17.15	&	18.37	&	-	&	Sch	&	FLI	\\
05.09.2012	&	56176.272	&	15.68	&	17.18	&	18.51	&	-	&	Sch	&	FLI	\\
08.09.2012	&	56179.272	&	15.57	&	17.09	&	-	&	-	&	60-cm	&	FLI	\\
09.09.2012	&	56180.364	&	15.51	&	16.99	&	18.06	&	-	&	60-cm	&	FLI	\\
22.09.2012	&	56193.339	&	15.59	&	17.05	&	18.32	&	-	&	Sch	&	FLI	\\
23.09.2012	&	56194.322	&	15.74	&	17.31	&	18.61	&	19.45	&	Sch	&	FLI	\\
07.10.2012	&	56208.210	&	15.79	&	17.31	&	18.66	&	-	&	Sch	&	FLI	\\
08.10.2012	&	56209.204	&	15.52	&	17.00	&	18.39	&	-	&	Sch	&	FLI	\\
09.10.2012	&	56210.193	&	15.59	&	17.13	&	18.54	&	-	&	Sch	&	FLI	\\
10.10.2012	&	56211.461	&	15.73	&	17.14	&	-	&	-	&	Sch	&	FLI	\\
25.10.2012	&	56226.240	&	15.65	&	17.18	&	18.60	&	-	&	Sch	&	FLI	\\
26.10.2012	&	56227.401	&	15.76	&	17.20	&	18.71	&	-	&	Sch	&	FLI	\\
30.10.2012	&	56231.322	&	16.21	&	17.50	&	-	&	-	&	60-cm	&	FLI	\\
17.11.2012	&	56249.172	&	16.12	&	17.81	&	19.26	&	-	&	Sch	&	FLI	\\
18.11.2012	&	56250.181	&	16.25	&	17.93	&	-	&	-	&	Sch	&	FLI	\\
31.12.2012	&	56293.254	&	15.53	&	17.05	&	18.31	&	-	&	Sch	&	FLI	\\
01.01.2013	&	56294.228	&	15.52	&	16.97	&	18.27	&	-	&	60-cm	&	FLI	\\
03.01.2013	&	56296.273	&	15.58	&	17.01	&	-	&	-	&	60-cm	&	FLI	\\
16.01.2013	&	56309.243	&	15.66	&	17.20	&	-	&	-	&	Sch	&	FLI	\\
04.02.2013	&	56328.205	&	15.55	&	-	&	-	&	-	&	Sch	&	FLI	\\
05.02.2013	&	56329.196	&	15.51	&	16.99	&	18.23	&	-	&	Sch	&	FLI	\\
06.03.2013	&	56357.595	&	15.31	&	16.65	&	17.68	&	-	&	60-cm	&	FLI	\\
10.04.2013	&	56392.510	&	15.17	&	16.49	&	-	&	-	&	Sch	&	FLI	\\
11.04.2013	&	56394.482	&	14.96	&	16.32	&	17.56	&	19.04	&	Sch	&	FLI	\\
02.05.2013	&	56415.418	&	15.45	&	16.87	&	18.12	&	19.38	&	Sch	&	FLI	\\
15.05.2013	&	56428.404	&	15.34	&	16.60	&	-	&	-	&	60-cm	&	FLI	\\
17.05.2013	&	56430.408	&	15.33	&	16.48	&	-	&	-	&	60-cm	&	FLI	\\
19.05.2013	&	56432.405	&	15.29	&	16.50	&	17.83	&	-	&	60-cm	&	FLI	\\
30.05.2013	&	56443.367	&	15.19	&	16.42	&	-	&	-	&	Sch	&	FLI	\\
31.05.2013	&	56444.354	&	15.35	&	16.68	&	18.08	&	-	&	Sch	&	FLI	\\
04.08.2013	&	56509.288	&	15.31	&	16.77	&	18.07	&	-	&	Sch	&	FLI	\\
05.08.2013	&	56510.369	&	15.51	&	16.87	&	18.38	&	-	&	60-cm	&	FLI	\\
05.08.2013	&	56510.386	&	15.35	&	16.78	&	18.11	&	-	&	Sch	&	FLI	\\
06.08.2013	&	56511.411	&	15.36	&	-	&	18.15	&	-	&	Sch	&	FLI	\\
06.08.2013	&	56511.413	&	15.46	&	16.75	&	18.36	&	-	&	60-cm	&	FLI	\\
07.08.2013	&	56512.398	&	15.36	&	16.82	&	18.17	&	-	&	Sch	&	FLI	\\
07.08.2013	&	56512.403	&	15.54	&	16.80	&	18.31	&	-	&	60-cm	&	FLI	\\
08.08.2013	&	56513.382	&	15.31	&	16.59	&	18.08	&	-	&	60-cm	&	FLI	\\
09.08.2013	&	56514.350	&	15.33	&	16.60	&	17.87	&	-	&	60-cm	&	FLI	\\
12.08.2013	&	56517.284	&	15.14	&	16.33	&	17.76	&	-	&	60-cm	&	FLI	\\
04.09.2013	&	56540.274	&	15.27	&	16.64	&	17.93	&	-	&	Sch	&	FLI	\\
05.09.2013	&	56541.322	&	15.20	&	16.64	&	17.90	&	-	&	Sch	&	FLI	\\
06.09.2013	&	56542.381	&	15.10	&	16.53	&	17.72	&	19.16	&	Sch	&	FLI	\\
11.09.2013	&	56547.383	&	15.06	&	16.22	&	-	&	-	&	60-cm	&	FLI	\\
14.09.2013	&	56550.363	&	15.07	&	16.22	&	17.44	&	18.60	&	60-cm	&	FLI	\\
11.10.2013	&	56577.306	&	15.27	&	-	&	17.93	&	-	&	60-cm	&	FLI	\\
12.10.2013	&	56578.331	&	15.26	&	16.56	&	17.85	&	19.05	&	60-cm	&	FLI	\\
07.11.2013	&	56604.264	&	15.12	&	16.35	&	17.61	&	-	&	60-cm	&	FLI	\\
28.12.2013	&	56655.200	&	15.22	&	16.51	&	17.91	&	19.46	&	Sch	&	FLI	\\
29.12.2013	&	56656.180	&	15.13	&	16.46	&	17.85	&	19.37	&	Sch	&	FLI	\\
30.12.2013	&	56657.193	&	15.03	&	16.36	&	17.71	&	19.24	&	Sch	&	FLI	\\
23.01.2014	&	56681.189	&	14.99	&	16.34	&	17.67	&	-	&	Sch	&	FLI	\\
22.03.2014	&	56738.590	&	15.08	&	16.46	&	17.80	&	19.37	&	Sch	&	FLI	\\
21.05.2014	&	56799.425	&	14.84	&	16.18	&	17.41	&	-	&	Sch	&	FLI	\\
28.06.2014	&	56837.417	&	14.96	&	16.29	&	17.45	&	18.99	&	Sch	&	FLI	\\
29.06.2014	&	56838.398	&	14.95	&	16.28	&	17.45	&	18.96	&	Sch	&	FLI	\\
20.07.2014	&	56859.390	&	15.01	&	16.23	&	17.15	&	-	&	60-cm	&	FLI	\\
21.07.2014	&	56860.391	&	15.06	&	16.35	&	17.28	&	-	&	60-cm	&	FLI	\\
24.07.2014	&	56863.339	&	14.92	&	16.30	&	17.57	&	-	&	Sch	&	FLI	\\
25.07.2014	&	56864.356	&	14.99	&	16.26	&	-	&	-	&	Sch	&	FLI	\\
03.08.2014	&	56873.386	&	15.02	&	16.43	&	17.73	&	19.40	&	Sch	&	FLI	\\
04.08.2014	&	56874.407	&	14.90	&	16.34	&	17.57	&	-	&	Sch	&	FLI	\\
18.08.2014	&	56888.450	&	14.92	&	16.28	&	17.53	&	19.09	&	Sch	&	FLI	\\
19.08.2014	&	56889.365	&	15.03	&	16.31	&	17.52	&	19.04	&	Sch	&	FLI	\\
26.11.2014	&	56988.181	&	15.21	&	16.63	&	17.90	&	-	&	Sch	&	FLI	\\
13.12.2014	&	57005.181	&	15.72	&	17.17	&	18.62	&	-	&	Sch	&	FLI	\\
14.12.2014	&	57006.220	&	15.61	&	17.02	&	18.41	&	-	&	Sch	&	FLI	\\
\hline \hline
\end{longtable}}

{\small
\begin{longtable}{cccccccc}
\caption{Photometric CCD observations and data of FHO 28 during the period June 1997$-$December 2014}\\
\hline\hline
\noalign{\smallskip}  
Date & J.D. (24...) & I & R & V & B & Tel & CCD\\
\noalign{\smallskip}  
\hline
\endfirsthead
\caption{Continued.}\\
\hline\hline
\noalign{\smallskip}  
Date & J.D. (24...) & I & R & V & B & Tel & CCD\\
\noalign{\smallskip}  
\hline
\noalign{\smallskip}  
\endhead
\hline
\label{Tab4}
\endfoot
\noalign{\smallskip}
01.06.1997	&	50601.491	&	15.81	&	17.34	&	18.96	&	-	&	2-m	&	Phot	\\
14.06.2000	&	51710.467	&	15.51	&	-	&	18.77	&	-	&	1.3-m	&	Phot	\\
16.06.2000	&	51711.507	&	15.52	&	-	&	18.78	&	-	&	1.3-m	&	Phot	\\
17.06.2000	&	51712.510	&	15.54	&	-	&	18.80	&	-	&	1.3-m	&	Phot	\\
20.06.2000	&	51716.377	&	16.65	&	-	&	20.23	&	-	&	1.3-m	&	Phot	\\
22.06.2000	&	51717.538	&	15.83	&	17.59	&	19.07	&	-	&	1.3-m	&	Phot	\\
23.06.2000	&	51718.545	&	15.58	&	-	&	18.79	&	-	&	1.3-m	&	Phot	\\
23.06.2000	&	51719.470	&	15.51	&	17.26	&	18.75	&	-	&	1.3-m	&	Phot	\\
25.06.2000	&	51720.515	&	15.52	&	17.28	&	18.74	&	-	&	1.3-m	&	Phot	\\
07.08.2000	&	51764.385	&	15.59	&	17.34	&	18.79	&	20.47	&	1.3-m	&	Phot	\\
29.10.2000	&	51847.241	&	15.65	&	17.52	&	19.25	&	-	&	Sch	&	ST-8	\\
30.10.2000	&	51848.300	&	15.58	&	17.43	&	18.88	&	-	&	Sch	&	ST-8	\\
24.12.2000	&	51903.243	&	16.06	&	18.39	&	-	&	-	&	Sch	&	ST-8	\\
27.05.2001	&	52057.425	&	15.81	&	17.87	&	-	&	-	&	Sch	&	ST-8	\\
16.06.2001	&	52106.549	&	15.62	&	-	&	18.90	&	-	&	1.3-m	&	Phot	\\
06.08.2001	&	52128.486	&	15.60	&	17.36	&	18.83	&	-	&	1.3-m	&	Phot	\\
02.09.2001	&	52154.515	&	15.85	&	-	&	19.21	&	-	&	1.3-m	&	Phot	\\
07.06.2002	&	52433.445	&	15.56	&	17.38	&	18.99	&	-	&	1.3-m	&	Phot	\\
23.06.2002	&	52449.000	&	15.66	&	-	&	19.05	&	-	&	1.3-m	&	Phot	\\
01.07.2002	&	52456.632	&	-	&	17.95	&	-	&	-	&	1.3-m	&	Phot	\\
15.07.2002	&	52471.468	&	15.49	&	-	&	18.72	&	-	&	1.3-m	&	Phot	\\
03.10.2002	&	52551.364	&	16.09	&	18.11	&	-	&	-	&	Sch	&	ST-8	\\
04.10.2002	&	52552.394	&	16.28	&	18.48	&	-	&	-	&	Sch	&	ST-8	\\
29.10.2002	&	52577.288	&	15.79	&	17.98	&	-	&	-	&	Sch	&	ST-8	\\
30.10.2002	&	52578.284	&	15.63	&	17.77	&	19.43	&	-	&	Sch	&	ST-8	\\
01.11.2002	&	52580.193	&	15.67	&	17.71	&	19.01	&	-	&	Sch	&	ST-8	\\
28.11.2002	&	52607.212	&	15.60	&	17.46	&	18.73	&	-	&	Sch	&	ST-8	\\
03.03.2003	&	52701.593	&	15.68	&	17.17	&	18.77	&	20.19	&	2-m	&	Phot	\\
06.05.2003	&	52765.517	&	15.99	&	17.91	&	19.40	&	-	&	Sch	&	ST-8	\\
27.09.2003	&	52910.375	&	15.67	&	17.44	&	18.70	&	-	&	Sch	&	ST-8	\\
25.11.2003	&	52969.192	&	15.66	&	17.72	&	18.82	&	-	&	Sch	&	ST-8	\\
22.03.2004	&	53086.582	&	15.66	&	17.32	&	18.73	&	-	&	2-m	&	Phot	\\
17.07.2004	&	53203.607	&	15.63	&	17.66	&	-	&	-	&	Sch	&	ST-8	\\
20.08.2004	&	53238.328	&	15.53	&	17.33	&	18.85	&	-	&	1.3-m	&	Phot	\\
08.09.2004	&	53257.324	&	15.49	&	17.29	&	18.75	&	-	&	1.3-m	&	Phot	\\
09.09.2004	&	53258.408	&	15.54	&	17.34	&	18.85	&	-	&	1.3-m	&	Phot	\\
28.09.2004	&	53277.227	&	15.95	&	17.88	&	19.25	&	-	&	1.3-m	&	Phot	\\
29.09.2004	&	53278.237	&	15.58	&	17.35	&	18.91	&	20.33	&	1.3-m	&	Phot	\\
30.09.2004	&	53279.274	&	15.90	&	17.76	&	19.31	&	-	&	1.3-m	&	Phot	\\
18.11.2004	&	53328.231	&	16.08	&	18.24	&	-	&	-	&	Sch	&	ST-8	\\
20.11.2004	&	53330.273	&	15.69	&	17.68	&	18.84	&	-	&	Sch	&	ST-8	\\
23.08.2005	&	53606.289	&	15.63	&	17.44	&	18.91	&	-	&	1.3-m	&	Phot	\\
27.08.2005	&	53609.509	&	15.71	&	17.58	&	19.07	&	-	&	1.3-m	&	Phot	\\
27.08.2005	&	53610.438	&	15.61	&	17.44	&	19.04	&	-	&	1.3-m	&	Phot	\\
14.09.2005	&	53628.281	&	15.63	&	17.44	&	18.95	&	-	&	1.3-m	&	Phot	\\
27.03.2006	&	53821.545	&	15.85	&	17.58	&	19.42	&	-	&	2-m	&	VA	\\
19.07.2006	&	53936.474	&	15.95	&	18.31	&	-	&	-	&	Sch	&	ST-8	\\
21.07.2006	&	53938.394	&	15.62	&	17.08	&	18.58	&	20.02	&	2-m	&	Phot	\\
30.09.2006	&	54009.385	&	15.64	&	17.47	&	18.89	&	-	&	1.3-m	&	Phot	\\
05.10.2006	&	54014.359	&	16.18	&	18.10	&	19.79	&	-	&	1.3-m	&	Phot	\\
16.12.2006	&	54086.204	&	16.79	&	18.84	&	-	&	-	&	Sch	&	ST-8	\\
26.06.2007	&	54278.380	&	15.53	&	17.27	&	18.87	&	-	&	1.3-m	&	Phot	\\
03.07.2007	&	54285.341	&	15.60	&	17.43	&	18.93	&	-	&	1.3-m	&	Phot	\\
23.07.2007	&	54305.319	&	16.01	&	17.97	&	19.64	&	-	&	1.3-m	&	ANDOR	\\
24.07.2007	&	54306.317	&	15.77	&	17.63	&	19.26	&	-	&	1.3-m	&	ANDOR	\\
16.08.2007	&	54329.384	&	15.43	&	17.00	&	18.65	&	-	&	2-m	&	VA	\\
17.08.2007	&	54330.289	&	15.44	&	17.02	&	18.68	&	-	&	2-m	&	VA	\\
28.06.2008	&	54646.366	&	15.75	&	17.71	&	19.35	&	-	&	1.3-m	&	ANDOR	\\
29.06.2008	&	54647.384	&	15.84	&	17.77	&	19.35	&	-	&	1.3-m	&	ANDOR	\\
05.07.2008	&	54653.347	&	15.67	&	17.45	&	18.93	&	-	&	1.3-m	&	ANDOR	\\
06.07.2008	&	54654.376	&	15.84	&	17.80	&	19.35	&	-	&	1.3-m	&	ANDOR	\\
25.07.2008	&	54673.331	&	16.06	&	18.03	&	19.70	&	-	&	1.3-m	&	ANDOR	\\
28.08.2008	&	54707.312	&	16.05	&	17.75	&	19.31	&	-	&	Sch	&	STL-11	\\
23.10.2008	&	54763.202	&	15.80	&	17.41	&	-	&	-	&	Sch	&	STL-11	\\
16.04.2009	&	54938.551	&	15.81	&	17.72	&	-	&	-	&	Sch	&	STL-11	\\
17.06.2009	&	55000.499	&	16.03	&	17.99	&	19.65	&	-	&	1.3-m	&	ANDOR	\\
27.06.2009	&	55009.522	&	15.69	&	17.56	&	19.05	&	-	&	1.3-m	&	ANDOR	\\
28.06.2009	&	55011.458	&	16.56	&	18.89	&	-	&	-	&	Sch	&	FLI	\\
04.07.2009	&	55016.501	&	15.95	&	17.75	&	19.31	&	-	&	1.3-m	&	ANDOR	\\
10.07.2009	&	55022.502	&	16.10	&	18.22	&	20.10	&	-	&	1.3-m	&	ANDOR	\\
14.07.2009	&	55027.412	&	16.13	&	18.32	&	19.30	&	-	&	Sch	&	FLI	\\
15.07.2009	&	55028.380	&	16.14	&	18.34	&	-	&	-	&	Sch	&	FLI	\\
31.07.2009	&	55044.338	&	15.89	&	17.79	&	19.30	&	-	&	1.3-m	&	ANDOR	\\
21.08.2009	&	55065.292	&	16.19	&	18.20	&	-	&	-	&	Sch	&	FLI	\\
06.10.2009	&	55111.330	&	16.02	&	-	&	-	&	-	&	Sch	&	FLI	\\
08.10.2009	&	55113.243	&	15.82	&	17.64	&	18.79	&	-	&	Sch	&	FLI	\\
20.11.2009	&	55156.197	&	16.13	&	18.24	&	-	&	-	&	Sch	&	FLI	\\
21.11.2009	&	55157.227	&	15.73	&	17.65	&	19.13	&	-	&	Sch	&	FLI	\\
13.05.2010	&	55330.436	&	15.83	&	17.78	&	-	&	-	&	Sch	&	FLI	\\
10.06.2010	&	55358.424	&	15.79	&	17.67	&	19.01	&	-	&	Sch	&	FLI	\\
06.08.2010	&	55415.466	&	15.74	&	17.62	&	18.96	&	-	&	Sch	&	FLI	\\
07.08.2010	&	55416.433	&	15.64	&	17.52	&	19.08	&	-	&	Sch	&	FLI	\\
11.08.2010	&	55420.497	&	15.69	&	17.66	&	19.20	&	-	&	1.3-m	&	ANDOR	\\
12.08.2010	&	55421.326	&	16.34	&	18.39	&	20.15	&	-	&	1.3-m	&	ANDOR	\\
13.08.2010	&	55422.258	&	15.85	&	17.76	&	19.32	&	-	&	1.3-m	&	ANDOR	\\
14.08.2010	&	55423.256	&	15.80	&	17.73	&	19.20	&	-	&	1.3-m	&	ANDOR	\\
16.08.2010	&	55425.250	&	15.62	&	17.38	&	18.91	&	20.39	&	1.3-m	&	ANDOR	\\
18.08.2010	&	55426.599	&	15.59	&	17.38	&	18.90	&	-	&	1.3-m	&	ANDOR	\\
19.08.2010	&	55427.591	&	15.60	&	17.41	&	18.99	&	-	&	1.3-m	&	ANDOR	\\
20.08.2010	&	55428.590	&	15.68	&	17.53	&	19.17	&	-	&	1.3-m	&	ANDOR	\\
21.08.2010	&	55429.589	&	15.79	&	17.71	&	19.28	&	-	&	1.3-m	&	ANDOR	\\
24.08.2010	&	55432.519	&	15.63	&	17.54	&	19.05	&	-	&	1.3-m	&	ANDOR	\\
24.08.2010	&	55433.470	&	15.67	&	17.49	&	19.11	&	-	&	1.3-m	&	ANDOR	\\
25.08.2010	&	55434.290	&	15.57	&	17.36	&	18.80	&	20.33	&	1.3-m	&	ANDOR	\\
26.08.2010	&	55435.310	&	15.57	&	17.39	&	18.82	&	20.24	&	1.3-m	&	ANDOR	\\
30.08.2010	&	55439.237	&	15.65	&	17.58	&	19.16	&	-	&	1.3-m	&	ANDOR	\\
31.08.2010	&	55440.238	&	15.68	&	17.45	&	18.93	&	-	&	1.3-m	&	ANDOR	\\
07.09.2010	&	55447.420	&	16.16	&	18.00	&	19.28	&	-	&	Sch	&	FLI	\\
08.09.2010	&	55448.326	&	15.75	&	17.61	&	18.93	&	-	&	Sch	&	FLI	\\
09.09.2010	&	55449.407	&	15.69	&	17.51	&	19.23	&	-	&	Sch	&	FLI	\\
18.09.2010	&	55458.218	&	15.82	&	17.70	&	19.29	&	20.47	&	1.3-m	&	ANDOR	\\
20.09.2010	&	55459.513	&	15.78	&	17.67	&	19.18	&	-	&	1.3-m	&	ANDOR	\\
11.10.2010	&	55481.276	&	15.72	&	17.62	&	19.12	&	-	&	1.3-m	&	ANDOR	\\
29.10.2010	&	55499.257	&	15.71	&	17.21	&	18.88	&	-	&	2-m	&	VA	\\
30.10.2010	&	55500.215	&	15.64	&	17.19	&	18.88	&	-	&	2-m	&	VA	\\
31.10.2010	&	55501.195	&	15.67	&	17.50	&	18.89	&	-	&	Sch	&	FLI	\\
31.10.2010	&	55501.287	&	15.60	&	17.15	&	18.78	&	-	&	2-m	&	VA	\\
01.11.2010	&	55502.234	&	16.12	&	17.84	&	19.66	&	-	&	2-m	&	VA	\\
02.11.2010	&	55503.223	&	16.99	&	-	&	-	&	-	&	Sch	&	FLI	\\
03.11.2010	&	55504.195	&	16.04	&	18.04	&	19.50	&	-	&	Sch	&	FLI	\\
04.11.2010	&	55505.202	&	16.03	&	18.06	&	19.29	&	-	&	Sch	&	FLI	\\
05.11.2010	&	55506.226	&	15.72	&	17.60	&	19.26	&	-	&	Sch	&	FLI	\\
06.11.2010	&	55507.224	&	15.72	&	17.63	&	19.38	&	-	&	Sch	&	FLI	\\
01.01.2011	&	55563.169	&	15.83	&	17.86	&	19.18	&	-	&	Sch	&	FLI	\\
06.01.2011	&	55568.198	&	15.70	&	17.15	&	18.73	&	20.44	&	2-m	&	VA	\\
07.01.2011	&	55569.210	&	15.56	&	17.06	&	18.59	&	-	&	2-m	&	VA	\\
08.01.2011	&	55570.204	&	15.48	&	17.00	&	18.62	&	20.47	&	2-m	&	VA	\\
09.01.2011	&	55571.216	&	15.70	&	17.30	&	19.03	&	-	&	2-m	&	VA	\\
10.01.2011	&	55572.175	&	-	&	17.22	&	-	&	-	&	2-m	&	VA	\\
11.01.2011	&	55573.232	&	16.32	&	18.02	&	19.87	&	-	&	2-m	&	VA	\\
12.01.2011	&	55574.219	&	15.84	&	17.27	&	18.85	&	-	&	2-m	&	VA	\\
07.02.2011	&	55600.202	&	16.60	&	-	&	-	&	-	&	Sch	&	FLI	\\
08.02.2011	&	55600.671	&	16.42	&	18.70	&	-	&	-	&	Sch	&	FLI	\\
04.04.2011	&	55656.509	&	15.83	&	17.88	&	-	&	-	&	Sch	&	FLI	\\
08.04.2011	&	55659.519	&	15.71	&	17.34	&	18.95	&	-	&	2-m	&	VA	\\
02.05.2011	&	55683.573	&	15.67	&	17.21	&	18.91	&	-	&	2-m	&	VA	\\
21.05.2011	&	55703.384	&	15.94	&	17.86	&	-	&	-	&	Sch	&	FLI	\\
22.05.2011	&	55704.430	&	15.79	&	17.70	&	-	&	-	&	Sch	&	FLI	\\
23.05.2011	&	55705.408	&	15.70	&	17.70	&	19.48	&	-	&	Sch	&	FLI	\\
24.05.2011	&	55706.382	&	16.36	&	-	&	-	&	-	&	Sch	&	FLI	\\
25.05.2011	&	55707.390	&	16.09	&	17.57	&	-	&	-	&	Sch	&	FLI	\\
08.06.2011	&	55721.380	&	15.96	&	17.41	&	19.17	&	-	&	2-m	&	VA	\\
09.06.2011	&	55722.425	&	15.73	&	17.59	&	-	&	-	&	Sch	&	FLI	\\
21.06.2011	&	55734.420	&	15.68	&	17.72	&	-	&	-	&	Sch	&	FLI	\\
22.06.2011	&	55735.452	&	15.72	&	17.57	&	19.16	&	-	&	Sch	&	FLI	\\
23.06.2011	&	55736.443	&	15.63	&	17.64	&	-	&	-	&	Sch	&	FLI	\\
24.06.2011	&	55737.443	&	15.78	&	17.81	&	-	&	-	&	Sch	&	FLI	\\
27.07.2011	&	55770.361	&	15.63	&	17.51	&	19.15	&	-	&	Sch	&	FLI	\\
16.08.2011	&	55790.386	&	15.71	&	17.60	&	19.15	&	-	&	1.3-m	&	ANDOR	\\
17.08.2011	&	55791.414	&	15.58	&	17.47	&	18.89	&	-	&	1.3-m	&	ANDOR	\\
18.08.2011	&	55792.397	&	15.92	&	17.82	&	19.38	&	-	&	1.3-m	&	ANDOR	\\
23.08.2011	&	55797.323	&	15.72	&	17.72	&	19.18	&	-	&	Sch	&	FLI	\\
24.08.2011	&	55798.301	&	15.76	&	17.77	&	19.27	&	-	&	Sch	&	FLI	\\
25.08.2011	&	55799.318	&	15.74	&	17.66	&	19.19	&	-	&	Sch	&	FLI	\\
10.09.2011	&	55815.249	&	15.61	&	17.55	&	19.10	&	-	&	1.3-m	&	ANDOR	\\
11.09.2011	&	55816.379	&	15.70	&	17.57	&	19.15	&	-	&	1.3-m	&	ANDOR	\\
19.09.2011	&	55824.281	&	15.61	&	17.59	&	19.07	&	-	&	1.3-m	&	ANDOR	\\
23.09.2011	&	55828.249	&	15.72	&	17.63	&	18.96	&	-	&	Sch	&	FLI	\\
07.10.2011	&	55842.269	&	15.59	&	17.54	&	19.11	&	-	&	1.3-m	&	ANDOR	\\
13.10.2011	&	55848.277	&	15.65	&	17.55	&	19.11	&	-	&	1.3-m	&	ANDOR	\\
29.10.2011	&	55864.251	&	15.64	&	17.16	&	18.82	&	-	&	2-m	&	VA	\\
30.10.2011	&	55865.186	&	15.66	&	17.16	&	18.93	&	-	&	2-m	&	VA	\\
31.10.2011	&	55866.282	&	15.69	&	17.29	&	19.06	&	-	&	2-m	&	VA	\\
26.11.2011	&	55892.170	&	15.63	&	17.12	&	18.78	&	-	&	2-m	&	VA	\\
27.11.2011	&	55893.176	&	15.98	&	18.16	&	-	&	-	&	Sch	&	FLI	\\
28.11.2011	&	55894.175	&	15.83	&	17.65	&	19.27	&	-	&	Sch	&	FLI	\\
29.11.2011	&	55895.167	&	15.87	&	17.83	&	19.46	&	-	&	Sch	&	FLI	\\
30.11.2011	&	55896.179	&	15.68	&	17.52	&	18.80	&	-	&	Sch	&	FLI	\\
29.12.2011	&	55925.169	&	15.80	&	17.50	&	19.01	&	-	&	Sch	&	FLI	\\
01.01.2012	&	55928.185	&	15.78	&	17.65	&	19.23	&	-	&	Sch	&	FLI	\\
31.01.2012	&	55958.188	&	15.97	&	17.47	&	-	&	-	&	2-m	&	VA	\\
16.03.2012	&	56003.563	&	15.94	&	17.80	&	-	&	-	&	Sch	&	FLI	\\
29.03.2012	&	56015.506	&	15.92	&	17.03	&	18.62	&	20.39	&	2-m	&	VA	\\
13.04.2012	&	56030.508	&	15.80	&	17.62	&	-	&	-	&	Sch	&	FLI	\\
12.05.2012	&	56060.415	&	16.22	&	18.11	&	-	&	-	&	Sch	&	FLI	\\
20.05.2012	&	56068.399	&	17.07	&	-	&	-	&	-	&	Sch	&	FLI	\\
12.06.2012	&	56091.398	&	16.24	&	18.15	&	19.37	&	-	&	Sch	&	FLI	\\
13.06.2012	&	56092.383	&	16.76	&	18.95	&	-	&	-	&	Sch	&	FLI	\\
15.06.2012	&	56094.441	&	16.43	&	18.07	&	19.80	&	-	&	2-m	&	VA	\\
17.06.2012	&	56096.398	&	16.23	&	18.22	&	-	&	-	&	Sch	&	FLI	\\
11.07.2012	&	56120.377	&	16.82	&	19.38	&	-	&	-	&	Sch	&	FLI	\\
12.07.2012	&	56121.338	&	16.26	&	18.24	&	-	&	-	&	Sch	&	FLI	\\
13.07.2012	&	56122.399	&	15.81	&	17.59	&	19.06	&	-	&	Sch	&	FLI	\\
14.07.2012	&	56123.387	&	15.71	&	17.44	&	18.80	&	-	&	Sch	&	FLI	\\
30.07.2012	&	56139.279	&	16.45	&	18.23	&	19.45	&	-	&	1.3-m	&	ANDOR	\\
01.08.2012	&	56141.416	&	16.01	&	17.71	&	19.23	&	-	&	1.3-m	&	ANDOR	\\
02.08.2012	&	56142.270	&	16.25	&	18.08	&	19.60	&	-	&	1.3-m	&	ANDOR	\\
03.08.2012	&	56143.254	&	16.01	&	17.97	&	19.38	&	-	&	1.3-m	&	ANDOR	\\
04.08.2012	&	56144.251	&	15.81	&	17.55	&	19.30	&	-	&	1.3-m	&	ANDOR	\\
11.08.2012	&	56150.615	&	15.84	&	17.67	&	19.14	&	-	&	1.3-m	&	ANDOR	\\
12.08.2012	&	56151.621	&	16.00	&	-	&	-	&	-	&	1.3-m	&	ANDOR	\\
13.08.2012	&	56152.624	&	16.12	&	18.00	&	-	&	-	&	1.3-m	&	ANDOR	\\
14.08.2012	&	56153.624	&	15.83	&	17.65	&	-	&	-	&	1.3-m	&	ANDOR	\\
15.08.2012	&	56154.622	&	15.91	&	17.93	&	19.93	&	-	&	1.3-m	&	ANDOR	\\
16.08.2012	&	56155.620	&	16.07	&	18.05	&	-	&	-	&	1.3-m	&	ANDOR	\\
16.08.2012	&	56156.242	&	15.85	&	-	&	-	&	-	&	1.3-m	&	ANDOR	\\
18.08.2012	&	56157.582	&	16.40	&	18.29	&	19.92	&	-	&	1.3-m	&	ANDOR	\\
19.08.2012	&	56159.351	&	15.75	&	17.54	&	18.75	&	-	&	Sch	&	FLI	\\
20.08.2012	&	56160.329	&	16.53	&	18.62	&	-	&	-	&	Sch	&	FLI	\\
21.08.2012	&	56160.529	&	16.36	&	18.35	&	19.97	&	-	&	1.3-m	&	ANDOR	\\
21.08.2012	&	56161.346	&	15.82	&	17.50	&	18.71	&	-	&	Sch	&	FLI	\\
22.08.2012	&	56162.338	&	15.69	&	17.59	&	18.87	&	-	&	Sch	&	FLI	\\
02.09.2012	&	56173.335	&	16.37	&	18.37	&	19.88	&	-	&	1.3-m	&	ANDOR	\\
03.09.2012	&	56174.306	&	16.01	&	17.98	&	19.57	&	-	&	1.3-m	&	ANDOR	\\
04.09.2012	&	56175.426	&	15.88	&	17.71	&	19.23	&	-	&	Sch	&	FLI	\\
05.09.2012	&	56176.272	&	15.76	&	17.73	&	19.00	&	-	&	Sch	&	FLI	\\
07.09.2012	&	56178.259	&	15.69	&	17.49	&	18.96	&	-	&	1.3-m	&	ANDOR	\\
08.09.2012	&	56179.272	&	15.74	&	17.55	&	18.94	&	-	&	60-cm	&	FLI	\\
08.09.2012	&	56179.467	&	15.66	&	17.52	&	19.00	&	-	&	1.3-m	&	ANDOR	\\
09.09.2012	&	56180.312	&	15.66	&	17.46	&	18.97	&	-	&	1.3-m	&	ANDOR	\\
09.09.2012	&	56180.364	&	15.74	&	17.69	&	-	&	-	&	60-cm	&	FLI	\\
10.09.2012	&	56181.288	&	15.62	&	17.46	&	18.97	&	20.45	&	1.3-m	&	ANDOR	\\
11.09.2012	&	56182.246	&	15.81	&	17.72	&	19.22	&	-	&	1.3-m	&	ANDOR	\\
12.09.2012	&	56183.370	&	16.11	&	18.13	&	19.81	&	-	&	1.3-m	&	ANDOR	\\
22.09.2012	&	56193.282	&	15.92	&	17.74	&	19.19	&	-	&	1.3-m	&	ANDOR	\\
22.09.2012	&	56193.339	&	15.98	&	17.79	&	19.17	&	-	&	Sch	&	FLI	\\
23.09.2012	&	56194.322	&	15.70	&	17.61	&	19.24	&	-	&	Sch	&	FLI	\\
07.10.2012	&	56208.210	&	16.22	&	18.06	&	-	&	-	&	Sch	&	FLI	\\
08.10.2012	&	56209.204	&	15.65	&	17.42	&	19.02	&	-	&	Sch	&	FLI	\\
09.10.2012	&	56210.193	&	15.63	&	17.49	&	18.97	&	-	&	Sch	&	FLI	\\
10.10.2012	&	56211.461	&	15.56	&	17.38	&	-	&	-	&	Sch	&	FLI	\\
13.10.2012	&	56214.193	&	15.70	&	17.19	&	18.89	&	20.48	&	2-m	&	VA	\\
25.10.2012	&	56226.240	&	15.69	&	17.59	&	19.01	&	-	&	Sch	&	FLI	\\
26.10.2012	&	56227.401	&	15.70	&	17.61	&	18.97	&	-	&	Sch	&	FLI	\\
30.10.2012	&	56231.322	&	15.66	&	17.45	&	-	&	-	&	60-cm	&	FLI	\\
17.11.2012	&	56249.172	&	15.82	&	17.65	&	19.32	&	-	&	Sch	&	FLI	\\
18.11.2012	&	56250.181	&	15.78	&	17.51	&	19.11	&	-	&	Sch	&	FLI	\\
12.12.2012	&	56274.163	&	15.49	&	16.94	&	18.60	&	-	&	2-m	&	VA	\\
13.12.2012	&	56275.215	&	15.76	&	17.24	&	18.92	&	-	&	2-m	&	VA	\\
14.12.2012	&	56276.165	&	15.94	&	17.47	&	19.26	&	-	&	2-m	&	VA	\\
15.12.2012	&	56277.161	&	15.54	&	17.35	&	18.78	&	-	&	2-m	&	VA	\\
31.12.2012	&	56293.254	&	15.72	&	17.60	&	19.41	&	-	&	Sch	&	FLI	\\
01.01.2013	&	56294.228	&	15.70	&	17.52	&	-	&	-	&	60-cm	&	FLI	\\
03.01.2013	&	56296.273	&	15.55	&	17.40	&	-	&	-	&	60-cm	&	FLI	\\
16.01.2013	&	56309.243	&	16.21	&	17.93	&	19.05	&	-	&	Sch	&	FLI	\\
19.01.2013	&	56312.225	&	15.62	&	16.98	&	18.55	&	-	&	2-m	&	VA	\\
04.02.2013	&	56328.205	&	15.86	&	17.16	&	-	&	-	&	Sch	&	FLI	\\
05.02.2013	&	56329.196	&	16.00	&	17.44	&	-	&	-	&	Sch	&	FLI	\\
06.03.2013	&	56357.595	&	15.87	&	17.77	&	-	&	-	&	60-cm	&	FLI	\\
18.03.2013	&	56369.610	&	16.19	&	17.76	&	19.36	&	-	&	2-m	&	VA	\\
10.04.2013	&	56392.510	&	15.92	&	17.99	&	-	&	-	&	Sch	&	FLI	\\
11.04.2013	&	56394.482	&	16.85	&	17.32	&	-	&	-	&	Sch	&	FLI	\\
02.05.2013	&	56415.418	&	15.60	&	16.89	&	18.77	&	-	&	Sch	&	FLI	\\
04.05.2013	&	56417.466	&	15.72	&	17.36	&	19.07	&	-	&	2-m	&	VA	\\
15.05.2013	&	56428.404	&	16.07	&	18.82	&	-	&	-	&	60-cm	&	FLI	\\
17.05.2013	&	56430.408	&	15.65	&	17.40	&	-	&	-	&	60-cm	&	FLI	\\
19.05.2013	&	56432.405	&	15.42	&	17.24	&	-	&	-	&	60-cm	&	FLI	\\
30.05.2013	&	56443.367	&	15.66	&	17.22	&	-	&	-	&	Sch	&	FLI	\\
31.05.2013	&	56444.354	&	16.04	&	17.83	&	-	&	-	&	Sch	&	FLI	\\
04.07.2013	&	56478.363	&	15.66	&	17.14	&	18.76	&	20.49	&	2-m	&	VA	\\
01.08.2013	&	56506.281	&	15.86	&	17.43	&	19.23	&	-	&	2-m	&	VA	\\
02.08.2013	&	56507.272	&	15.73	&	17.21	&	18.93	&	-	&	2-m	&	VA	\\
03.08.2013	&	56508.318	&	15.85	&	17.42	&	19.15	&	-	&	2-m	&	VA	\\
04.08.2013	&	56509.288	&	15.84	&	17.67	&	19.33	&	-	&	Sch	&	FLI	\\
05.08.2013	&	56510.369	&	15.53	&	17.26	&	18.78	&	-	&	60-cm	&	FLI	\\
05.08.2013	&	56510.386	&	15.56	&	17.38	&	18.84	&	-	&	Sch	&	FLI	\\
06.08.2013	&	56511.411	&	15.58	&	17.34	&	18.72	&	-	&	Sch	&	FLI	\\
06.08.2013	&	56511.413	&	15.51	&	17.25	&	18.75	&	-	&	60-cm	&	FLI	\\
07.08.2013	&	56512.398	&	15.66	&	17.46	&	18.93	&	-	&	Sch	&	FLI	\\
07.08.2013	&	56512.403	&	15.63	&	17.38	&	-	&	-	&	60-cm	&	FLI	\\
08.08.2013	&	56513.382	&	15.52	&	17.17	&	18.77	&	-	&	60-cm	&	FLI	\\
09.08.2013	&	56514.350	&	15.73	&	17.73	&	-	&	-	&	60-cm	&	FLI	\\
12.08.2013	&	56517.284	&	15.55	&	17.48	&	18.70	&	-	&	60-cm	&	FLI	\\
04.09.2013	&	56540.274	&	15.71	&	17.51	&	19.02	&	-	&	Sch	&	FLI	\\
05.09.2013	&	56541.322	&	15.72	&	17.56	&	19.11	&	-	&	Sch	&	FLI	\\
06.09.2013	&	56542.381	&	15.93	&	17.88	&	19.40	&	-	&	Sch	&	FLI	\\
07.09.2013	&	56543.396	&	15.69	&	17.05	&	18.64	&	20.39	&	2-m	&	VA	\\
08.09.2013	&	56544.264	&	15.70	&	17.19	&	18.82	&	20.47	&	2-m	&	VA	\\
11.09.2013	&	56547.383	&	15.55	&	17.41	&	-	&	-	&	60-cm	&	FLI	\\
14.09.2013	&	56550.363	&	15.69	&	17.57	&	-	&	-	&	60-cm	&	FLI	\\
17.09.2013	&	56553.282	&	15.47	&	17.26	&	18.79	&	20.27	&	1.3-m	&	ANDOR	\\
11.10.2013	&	56577.306	&	15.65	&	17.38	&	18.65	&	-	&	60-cm	&	FLI	\\
12.10.2013	&	56578.331	&	15.49	&	17.20	&	-	&	-	&	60-cm	&	FLI	\\
07.11.2013	&	56604.264	&	15.67	&	17.45	&	18.77	&	-	&	60-cm	&	FLI	\\
09.12.2013	&	56636.185	&	15.98	&	17.55	&	19.26	&	-	&	2-m	&	VA	\\
28.12.2013	&	56655.200	&	15.52	&	17.31	&	18.62	&	-	&	Sch	&	FLI	\\
29.12.2013	&	56656.180	&	15.66	&	17.42	&	18.96	&	-	&	Sch	&	FLI	\\
30.12.2013	&	56657.193	&	15.82	&	17.68	&	19.33	&	-	&	Sch	&	FLI	\\
23.01.2014	&	56681.189	&	15.60	&	17.00	&	-	&	-	&	Sch	&	FLI	\\
05.02.2014	&	56694.194	&	-	&	18.78	&	-	&	-	&	2-m	&	VA	\\
22.03.2014	&	56738.590	&	16.77	&	19.07	&	-	&	-	&	Sch	&	FLI	\\
21.05.2014	&	56799.425	&	15.80	&	17.66	&	-	&	-	&	Sch	&	FLI	\\
23.05.2014	&	56801.436	&	16.02	&	17.63	&	19.28	&	-	&	2-m	&	VA	\\
23.06.2014	&	56832.391	&	15.99	&	17.70	&	19.52	&	-	&	2-m	&	VA	\\
25.06.2014	&	56834.357	&	15.72	&	17.24	&	18.92	&	-	&	2-m	&	VA	\\
26.06.2014	&	56835.462	&	15.70	&	17.22	&	18.86	&	-	&	2-m	&	VA	\\
28.06.2014	&	56837.417	&	15.83	&	17.70	&	-	&	-	&	Sch	&	FLI	\\
29.06.2014	&	56838.398	&	15.77	&	17.61	&	19.21	&	-	&	Sch	&	FLI	\\
20.07.2014	&	56859.390	&	15.54	&	17.27	&	-	&	-	&	60-cm	&	FLI	\\
21.07.2014	&	56860.391	&	15.68	&	17.77	&	-	&	-	&	60-cm	&	FLI	\\
24.07.2014	&	56863.339	&	15.64	&	17.49	&	-	&	-	&	Sch	&	FLI	\\
25.07.2014	&	56864.356	&	15.68	&	17.34	&	-	&	-	&	Sch	&	FLI	\\
03.08.2014	&	56873.386	&	15.74	&	17.80	&	18.92	&	-	&	Sch	&	FLI	\\
04.08.2014	&	56874.407	&	15.77	&	17.76	&	-	&	-	&	Sch	&	FLI	\\
18.08.2014	&	56888.450	&	15.74	&	17.74	&	18.99	&	-	&	Sch	&	FLI	\\
19.08.2014	&	56889.365	&	16.00	&	17.99	&	19.26	&	-	&	Sch	&	FLI	\\
29.08.2014	&	56899.303	&	16.62	&	18.73	&	-	&	-	&	1.3-m	&	ANDOR	\\
26.11.2014	&	56988.181	&	15.67	&	17.53	&	18.86	&	-	&	Sch	&	FLI	\\
13.12.2014	&	57005.181	&	15.55	&	17.36	&	18.90	&	-	&	Sch	&	FLI	\\
14.12.2014	&	57006.220	&	15.73	&	17.50	&	19.01	&	-	&	Sch	&	FLI	\\
24.12.2014	&	57016.174	&	15.86	&	17.52	&	19.13	&	-	&	2-m	&	VA	\\
25.12.2014	&	57017.175	&	15.88	&	17.55	&	19.30	&	-	&	2-m	&	VA	\\
\hline \hline
\end{longtable}}

{\small
\begin{longtable}{cccccccc}
\caption{Photometric CCD observations and data of FHO 29 during the period November 2004$-$December 2014}\\
\hline\hline
\noalign{\smallskip}  
Date & J.D. (24...) & I & R & V & B & Tel & CCD\\
\noalign{\smallskip}  
\hline
\endfirsthead
\caption{Continued.}\\
\hline\hline
\noalign{\smallskip}  
Date & J.D. (24...) & I & R & V & B & Tel & CCD\\
\noalign{\smallskip}  
\hline
\noalign{\smallskip}  
\endhead
\hline
\label{Tab5}
\endfoot
\noalign{\smallskip}
20.11.2004	&	53330.273	&	14.77	&	15.84	&	16.85	&	17.96	&	Sch	&	ST-8	\\
16.12.2006	&	54086.204	&	15.83	&	16.83	&	-	&	19.00	&	Sch	&	ST-8	\\
28.08.2008	&	54707.312	&	15.57	&	16.48	&	17.53	&	18.74	&	Sch	&	STL-11	\\
16.04.2009	&	54938.551	&	16.19	&	17.21	&	-	&	-	&	Sch	&	STL-11	\\
28.06.2009	&	55011.458	&	14.93	&	15.95	&	16.90	&	-	&	Sch	&	FLI	\\
14.07.2009	&	55027.412	&	15.49	&	16.67	&	17.71	&	-	&	Sch	&	FLI	\\
15.07.2009	&	55028.380	&	15.65	&	16.84	&	18.01	&	-	&	Sch	&	FLI	\\
21.08.2009	&	55065.292	&	15.73	&	16.85	&	17.95	&	-	&	Sch	&	FLI	\\
06.10.2009	&	55111.330	&	16.38	&	-	&	17.84	&	19.16	&	Sch	&	FLI	\\
08.10.2009	&	55113.243	&	16.56	&	17.21	&	18.12	&	19.32	&	Sch	&	FLI	\\
20.11.2009	&	55156.197	&	16.61	&	17.22	&	18.06	&	19.24	&	Sch	&	FLI	\\
13.05.2010	&	55330.436	&	15.37	&	16.43	&	17.33	&	18.46	&	Sch	&	FLI	\\
10.06.2010	&	55358.424	&	14.29	&	15.28	&	16.27	&	17.48	&	Sch	&	FLI	\\
06.08.2010	&	55415.466	&	15.94	&	17.20	&	18.31	&	-	&	Sch	&	FLI	\\
07.08.2010	&	55416.433	&	15.71	&	16.92	&	18.12	&	-	&	Sch	&	FLI	\\
07.09.2010	&	55447.420	&	15.90	&	16.98	&	18.08	&	19.32	&	Sch	&	FLI	\\
08.09.2010	&	55448.326	&	15.60	&	16.77	&	17.83	&	19.05	&	Sch	&	FLI	\\
09.09.2010	&	55449.407	&	15.76	&	16.82	&	17.85	&	19.10	&	Sch	&	FLI	\\
31.10.2010	&	55501.195	&	16.50	&	17.43	&	18.28	&	19.33	&	Sch	&	FLI	\\
02.11.2010	&	55503.223	&	16.78	&	17.65	&	18.74	&	-	&	Sch	&	FLI	\\
03.11.2010	&	55504.195	&	16.56	&	17.31	&	18.36	&	19.33	&	Sch	&	FLI	\\
04.11.2010	&	55505.202	&	16.22	&	17.13	&	18.02	&	19.20	&	Sch	&	FLI	\\
05.11.2010	&	55506.226	&	16.21	&	17.09	&	18.13	&	19.20	&	Sch	&	FLI	\\
06.11.2010	&	55507.224	&	15.83	&	16.90	&	17.89	&	18.86	&	Sch	&	FLI	\\
01.01.2011	&	55563.169	&	15.60	&	16.58	&	17.61	&	18.65	&	Sch	&	FLI	\\
07.02.2011	&	55600.202	&	14.60	&	15.67	&	16.78	&	-	&	Sch	&	FLI	\\
08.02.2011	&	55600.671	&	14.57	&	15.62	&	16.66	&	-	&	Sch	&	FLI	\\
04.04.2011	&	55656.509	&	16.25	&	17.10	&	17.79	&	-	&	Sch	&	FLI	\\
21.05.2011	&	55703.384	&	16.68	&	17.42	&	18.14	&	-	&	Sch	&	FLI	\\
22.05.2011	&	55704.430	&	16.37	&	16.98	&	17.64	&	-	&	Sch	&	FLI	\\
23.05.2011	&	55705.408	&	16.31	&	16.97	&	17.62	&	18.56	&	Sch	&	FLI	\\
24.05.2011	&	55706.382	&	16.33	&	17.04	&	17.75	&	-	&	Sch	&	FLI	\\
25.05.2011	&	55707.390	&	16.47	&	17.29	&	17.89	&	-	&	Sch	&	FLI	\\
09.06.2011	&	55722.425	&	16.74	&	17.43	&	18.14	&	19.23	&	Sch	&	FLI	\\
21.06.2011	&	55734.420	&	16.15	&	16.96	&	17.88	&	19.07	&	Sch	&	FLI	\\
22.06.2011	&	55735.452	&	16.11	&	17.00	&	17.85	&	18.98	&	Sch	&	FLI	\\
23.06.2011	&	55736.443	&	16.29	&	17.32	&	18.09	&	-	&	Sch	&	FLI	\\
24.06.2011	&	55737.443	&	16.16	&	17.08	&	17.90	&	18.92	&	Sch	&	FLI	\\
27.07.2011	&	55770.361	&	15.61	&	16.64	&	17.54	&	18.58	&	Sch	&	FLI	\\
23.08.2011	&	55797.323	&	15.79	&	16.84	&	17.88	&	18.90	&	Sch	&	FLI	\\
24.08.2011	&	55798.301	&	15.78	&	16.80	&	17.78	&	18.79	&	Sch	&	FLI	\\
25.08.2011	&	55799.318	&	15.75	&	16.76	&	17.72	&	18.93	&	Sch	&	FLI	\\
23.09.2011	&	55828.249	&	15.74	&	16.83	&	17.90	&	19.23	&	Sch	&	FLI	\\
27.11.2011	&	55893.176	&	14.74	&	15.68	&	16.54	&	17.74	&	Sch	&	FLI	\\
28.11.2011	&	55894.175	&	15.00	&	16.03	&	16.99	&	18.14	&	Sch	&	FLI	\\
29.11.2011	&	55895.167	&	15.05	&	16.12	&	17.11	&	18.31	&	Sch	&	FLI	\\
30.11.2011	&	55896.179	&	14.84	&	15.75	&	16.67	&	17.74	&	Sch	&	FLI	\\
29.12.2011	&	55925.169	&	15.38	&	16.41	&	17.25	&	18.28	&	Sch	&	FLI	\\
01.01.2012	&	55928.185	&	16.14	&	16.95	&	17.82	&	18.85	&	Sch	&	FLI	\\
16.03.2012	&	56003.563	&	14.65	&	15.75	&	16.84	&	18.27	&	Sch	&	FLI	\\
13.04.2012	&	56030.508	&	15.58	&	16.67	&	17.61	&	19.23	&	Sch	&	FLI	\\
12.05.2012	&	56060.415	&	15.27	&	16.48	&	17.59	&	-	&	Sch	&	FLI	\\
20.05.2012	&	56068.399	&	14.34	&	15.45	&	16.37	&	-	&	Sch	&	FLI	\\
12.06.2012	&	56091.398	&	15.10	&	16.24	&	17.29	&	18.43	&	Sch	&	FLI	\\
13.06.2012	&	56092.383	&	15.12	&	16.23	&	17.27	&	18.39	&	Sch	&	FLI	\\
17.06.2012	&	56096.398	&	14.48	&	15.56	&	16.62	&	17.92	&	Sch	&	FLI	\\
11.07.2012	&	56120.377	&	15.02	&	16.17	&	17.17	&	18.59	&	Sch	&	FLI	\\
12.07.2012	&	56121.338	&	14.46	&	15.58	&	16.59	&	17.94	&	Sch	&	FLI	\\
13.07.2012	&	56122.399	&	14.81	&	15.92	&	17.01	&	18.33	&	Sch	&	FLI	\\
14.07.2012	&	56123.387	&	15.45	&	16.51	&	17.62	&	18.93	&	Sch	&	FLI	\\
19.08.2012	&	56159.351	&	15.63	&	16.72	&	17.76	&	18.79	&	Sch	&	FLI	\\
20.08.2012	&	56160.329	&	15.52	&	16.56	&	17.50	&	18.67	&	Sch	&	FLI	\\
21.08.2012	&	56161.346	&	15.86	&	16.70	&	17.71	&	18.83	&	Sch	&	FLI	\\
22.08.2012	&	56162.338	&	16.08	&	16.78	&	17.66	&	18.64	&	Sch	&	FLI	\\
04.09.2012	&	56175.426	&	15.24	&	16.35	&	17.27	&	18.00	&	Sch	&	FLI	\\
05.09.2012	&	56176.272	&	15.41	&	16.42	&	17.36	&	18.45	&	Sch	&	FLI	\\
08.09.2012	&	56179.272	&	16.95	&	17.72	&	18.53	&	-	&	60-cm	&	FLI	\\
09.09.2012	&	56180.364	&	16.61	&	17.28	&	-	&	-	&	60-cm	&	FLI	\\
22.09.2012	&	56193.339	&	16.65	&	17.65	&	18.57	&	-	&	Sch	&	FLI	\\
23.09.2012	&	56194.322	&	16.63	&	17.42	&	18.13	&	19.43	&	Sch	&	FLI	\\
07.10.2012	&	56208.210	&	16.39	&	17.00	&	17.78	&	18.80	&	Sch	&	FLI	\\
08.10.2012	&	56209.204	&	16.69	&	17.51	&	18.37	&	19.23	&	Sch	&	FLI	\\
09.10.2012	&	56210.193	&	16.85	&	17.59	&	18.43	&	19.43	&	Sch	&	FLI	\\
10.10.2012	&	56211.461	&	16.78	&	17.73	&	-	&	-	&	Sch	&	FLI	\\
25.10.2012	&	56226.240	&	16.63	&	17.38	&	18.14	&	18.99	&	Sch	&	FLI	\\
26.10.2012	&	56227.401	&	17.15	&	17.67	&	18.61	&	-	&	Sch	&	FLI	\\
30.10.2012	&	56231.322	&	16.85	&	-	&	-	&	-	&	60-cm	&	FLI	\\
17.11.2012	&	56249.172	&	16.80	&	17.53	&	18.36	&	19.48	&	Sch	&	FLI	\\
18.11.2012	&	56250.181	&	16.80	&	17.48	&	18.32	&	19.32	&	Sch	&	FLI	\\
31.12.2012	&	56293.254	&	15.48	&	16.58	&	17.63	&	18.81	&	Sch	&	FLI	\\
01.01.2013	&	56294.228	&	15.52	&	16.54	&	17.66	&	-	&	60-cm	&	FLI	\\
03.01.2013	&	56296.273	&	16.07	&	16.96	&	-	&	-	&	60-cm	&	FLI	\\
16.01.2013	&	56309.243	&	15.46	&	16.47	&	17.54	&	18.46	&	Sch	&	FLI	\\
05.02.2013	&	56329.196	&	16.25	&	17.27	&	17.84	&	-	&	Sch	&	FLI	\\
06.03.2013	&	56357.595	&	16.93	&	17.77	&	-	&	-	&	60-cm	&	FLI	\\
11.04.2013	&	56394.482	&	17.13	&	17.91	&	18.50	&	-	&	Sch	&	FLI	\\
02.05.2013	&	56415.418	&	17.47	&	18.12	&	19.01	&	-	&	Sch	&	FLI	\\
15.05.2013	&	56428.404	&	16.18	&	17.25	&	-	&	-	&	60-cm	&	FLI	\\
17.05.2013	&	56430.408	&	15.26	&	16.47	&	-	&	-	&	60-cm	&	FLI	\\
19.05.2013	&	56432.405	&	15.89	&	17.16	&	17.78	&	-	&	60-cm	&	FLI	\\
30.05.2013	&	56443.367	&	16.37	&	17.17	&	18.22	&	-	&	Sch	&	FLI	\\
31.05.2013	&	56444.354	&	16.53	&	17.29	&	18.22	&	-	&	Sch	&	FLI	\\
04.08.2013	&	56509.288	&	16.29	&	17.31	&	18.30	&	19.21	&	Sch	&	FLI	\\
05.08.2013	&	56510.369	&	16.26	&	17.56	&	18.58	&	-	&	60-cm	&	FLI	\\
05.08.2013	&	56510.386	&	16.36	&	17.40	&	18.54	&	19.43	&	Sch	&	FLI	\\
06.08.2013	&	56511.411	&	16.30	&	17.33	&	18.35	&	-	&	Sch	&	FLI	\\
06.08.2013	&	56511.413	&	16.37	&	17.32	&	18.29	&	19.42	&	60-cm	&	FLI	\\
07.08.2013	&	56512.398	&	15.83	&	17.01	&	17.90	&	19.10	&	Sch	&	FLI	\\
07.08.2013	&	56512.403	&	15.89	&	17.01	&	17.93	&	-	&	60-cm	&	FLI	\\
08.08.2013	&	56513.382	&	15.78	&	16.97	&	17.99	&	18.92	&	60-cm	&	FLI	\\
09.08.2013	&	56514.350	&	16.76	&	17.83	&	-	&	-	&	60-cm	&	FLI	\\
12.08.2013	&	56517.284	&	16.09	&	17.12	&	18.14	&	-	&	60-cm	&	FLI	\\
04.09.2013	&	56540.274	&	15.34	&	16.50	&	17.49	&	18.75	&	Sch	&	FLI	\\
05.09.2013	&	56541.322	&	15.13	&	16.24	&	17.29	&	18.45	&	Sch	&	FLI	\\
06.09.2013	&	56542.381	&	14.95	&	16.11	&	17.16	&	18.39	&	Sch	&	FLI	\\
11.09.2013	&	56547.383	&	15.60	&	16.76	&	17.74	&	-	&	60-cm	&	FLI	\\
14.09.2013	&	56550.363	&	14.64	&	15.58	&	16.47	&	17.53	&	60-cm	&	FLI	\\
11.10.2013	&	56577.306	&	15.51	&	16.58	&	17.58	&	-	&	60-cm	&	FLI	\\
12.10.2013	&	56578.331	&	15.14	&	16.15	&	17.16	&	18.26	&	60-cm	&	FLI	\\
07.11.2013	&	56604.264	&	-	&	17.74	&	18.91	&	-	&	60-cm	&	FLI	\\
28.12.2013	&	56655.200	&	16.28	&	17.09	&	17.90	&	19.22	&	Sch	&	FLI	\\
29.12.2013	&	56656.180	&	16.02	&	16.97	&	17.94	&	19.15	&	Sch	&	FLI	\\
30.12.2013	&	56657.193	&	15.52	&	16.72	&	17.88	&	18.93	&	Sch	&	FLI	\\
23.01.2014	&	56681.189	&	16.33	&	17.00	&	-	&	-	&	Sch	&	FLI	\\
22.03.2014	&	56738.590	&	16.69	&	17.52	&	18.17	&	19.02	&	Sch	&	FLI	\\
21.05.2014	&	56799.425	&	16.66	&	17.52	&	-	&	-	&	Sch	&	FLI	\\
28.06.2014	&	56837.417	&	16.15	&	17.32	&	18.43	&	-	&	Sch	&	FLI	\\
29.06.2014	&	56838.398	&	16.04	&	17.25	&	18.22	&	-	&	Sch	&	FLI	\\
21.07.2014	&	56860.391	&	15.66	&	16.82	&	17.70	&	-	&	60-cm	&	FLI	\\
24.07.2014	&	56863.339	&	15.51	&	16.66	&	-	&	19.03	&	Sch	&	FLI	\\
25.07.2014	&	56864.356	&	15.65	&	16.67	&	-	&	-	&	Sch	&	FLI	\\
03.08.2014	&	56873.386	&	15.89	&	16.64	&	17.44	&	18.39	&	Sch	&	FLI	\\
04.08.2014	&	56874.407	&	16.01	&	16.74	&	17.54	&	18.75	&	Sch	&	FLI	\\
18.08.2014	&	56888.450	&	15.62	&	16.71	&	17.73	&	19.26	&	Sch	&	FLI	\\
19.08.2014	&	56889.365	&	15.62	&	16.80	&	17.87	&	19.41	&	Sch	&	FLI	\\
26.11.2014	&	56988.181	&	15.55	&	16.55	&	17.54	&	18.71	&	Sch	&	FLI	\\
13.12.2014	&	57005.181	&	15.04	&	16.19	&	17.03	&	18.34	&	Sch	&	FLI	\\
14.12.2014	&	57006.220	&	15.24	&	16.38	&	17.26	&	-	&	Sch	&	FLI	\\
\hline \hline
\end{longtable}}

{\small
\begin{longtable}{cccccccc}
\caption{Photometric CCD observations and data of V1929 Cygni during the period October 2010$-$December 2014}\\
\hline\hline
\noalign{\smallskip}  
Date & J.D. (24...) & I & R & V & B & Tel & CCD\\
\noalign{\smallskip}  
\hline
\endfirsthead
\caption{Continued.}\\
\hline\hline
\noalign{\smallskip}  
Date & J.D. (24...) & I & R & V & B & Tel & CCD\\
\noalign{\smallskip}  
\hline
\noalign{\smallskip}  
\endhead
\hline
\label{Tab6}
\endfoot
\noalign{\smallskip}
29.10.2000	&	51847.241	&	13.85	&	14.47	&	15.16	&	16.16	&	Sch	&	ST-8	\\
30.10.2000	&	51848.300	&	13.93	&	14.60	&	15.27	&	16.26	&	Sch	&	ST-8	\\
24.12.2000	&	51903.243	&	13.87	&	14.49	&	15.19	&	-	&	Sch	&	ST-8	\\
27.05.2001	&	52057.425	&	13.98	&	14.62	&	15.27	&	-	&	Sch	&	ST-8	\\
03.10.2002	&	52551.364	&	13.83	&	14.42	&	15.06	&	16.01	&	Sch	&	ST-8	\\
04.10.2002	&	52552.394	&	13.85	&	14.43	&	15.07	&	16.00	&	Sch	&	ST-8	\\
29.10.2002	&	52577.288	&	13.93	&	14.60	&	15.25	&	16.26	&	Sch	&	ST-8	\\
30.10.2002	&	52578.284	&	13.84	&	14.41	&	15.06	&	16.00	&	Sch	&	ST-8	\\
01.11.2002	&	52580.193	&	13.85	&	14.48	&	15.14	&	16.12	&	Sch	&	ST-8	\\
28.11.2008	&	52607.212	&	13.79	&	14.40	&	15.03	&	15.91	&	Sch	&	ST-8	\\
27.09.2003	&	52910.375	&	13.85	&	14.50	&	15.18	&	16.16	&	Sch	&	ST-8	\\
25.11.2003	&	52969.192	&	13.83	&	14.50	&	15.15	&	16.15	&	Sch	&	ST-8	\\
17.07.2004	&	53203.607	&	13.88	&	14.51	&	15.18	&	-	&	Sch	&	ST-8	\\
18.11.2004	&	53328.231	&	13.85	&	14.47	&	15.11	&	16.12	&	Sch	&	ST-8	\\
20.11.2004	&	53330.273	&	13.78	&	14.38	&	15.03	&	16.00	&	Sch	&	ST-8	\\
16.07.2006	&	53936.474	&	13.73	&	14.31	&	14.95	&	-	&	Sch	&	ST-8	\\
28.08.2008	&	54707.312	&	13.64	&	14.22	&	14.84	&	15.77	&	Sch	&	STL-11	\\
23.10.2008	&	54763.202	&	13.68	&	14.27	&	14.90	&	15.97	&	Sch	&	STL-11	\\
16.04.2009	&	54938.551	&	13.65	&	14.20	&	-	&	-	&	Sch	&	STL-11	\\
28.06.2009	&	55011.458	&	13.67	&	14.23	&	14.83	&	-	&	Sch	&	FLI	\\
14.07.2009	&	55027.412	&	13.70	&	14.29	&	14.91	&	-	&	Sch	&	FLI	\\
15.07.2009	&	55028.380	&	13.62	&	14.19	&	14.80	&	15.82	&	Sch	&	FLI	\\
21.08.2009	&	55065.292	&	13.66	&	14.25	&	14.86	&	15.89	&	Sch	&	FLI	\\
06.10.2009	&	55111.330	&	13.69	&	-	&	14.89	&	15.92	&	Sch	&	FLI	\\
08.10.2009	&	55113.243	&	13.66	&	14.26	&	14.89	&	15.93	&	Sch	&	FLI	\\
20.11.2009	&	55156.197	&	13.66	&	14.23	&	14.83	&	15.81	&	Sch	&	FLI	\\
21.11.2009	&	55157.227	&	13.67	&	14.25	&	14.86	&	15.87	&	Sch	&	FLI	\\
13.05.2010	&	55330.436	&	13.66	&	14.26	&	14.89	&	15.90	&	Sch	&	FLI	\\
10.06.2010	&	55358.424	&	13.67	&	14.26	&	14.89	&	15.90	&	Sch	&	FLI	\\
06.08.2010	&	55415.466	&	13.67	&	14.29	&	14.91	&	15.90	&	Sch	&	FLI	\\
07.08.2010	&	55416.433	&	13.73	&	14.33	&	15.00	&	16.02	&	Sch	&	FLI	\\
07.09.2010	&	55447.420	&	13.66	&	14.26	&	14.87	&	15.86	&	Sch	&	FLI	\\
08.09.2010	&	55448.326	&	13.69	&	14.30	&	14.90	&	15.89	&	Sch	&	FLI	\\
09.09.2010	&	55449.407	&	13.74	&	14.36	&	14.99	&	16.01	&	Sch	&	FLI	\\
31.10.2010	&	55501.195	&	13.72	&	14.33	&	14.96	&	15.94	&	Sch	&	FLI	\\
02.11.2010	&	55503.223	&	13.73	&	14.36	&	14.96	&	15.97	&	Sch	&	FLI	\\
03.11.2010	&	55504.195	&	13.75	&	14.34	&	14.97	&	15.98	&	Sch	&	FLI	\\
04.11.2010	&	55505.202	&	13.69	&	14.31	&	14.94	&	15.93	&	Sch	&	FLI	\\
05.11.2010	&	55506.226	&	13.72	&	14.34	&	14.94	&	15.94	&	Sch	&	FLI	\\
06.11.2010	&	55507.224	&	13.77	&	14.38	&	15.01	&	16.04	&	Sch	&	FLI	\\
01.01.2011	&	55563.169	&	13.72	&	14.32	&	14.93	&	15.96	&	Sch	&	FLI	\\
07.02.2011	&	55600.202	&	13.76	&	14.38	&	15.04	&	-	&	Sch	&	FLI	\\
08.02.2011	&	55600.671	&	13.71	&	14.31	&	14.91	&	-	&	Sch	&	FLI	\\
04.04.2011	&	55656.509	&	13.67	&	14.26	&	14.85	&	15.83	&	Sch	&	FLI	\\
21.05.2011	&	55703.384	&	13.67	&	14.25	&	14.82	&	15.85	&	Sch	&	FLI	\\
22.05.2011	&	55704.430	&	13.71	&	14.31	&	14.92	&	15.95	&	Sch	&	FLI	\\
23.05.2011	&	55705.408	&	13.77	&	14.41	&	15.04	&	16.09	&	Sch	&	FLI	\\
24.05.2011	&	55706.382	&	13.65	&	14.23	&	14.82	&	15.82	&	Sch	&	FLI	\\
25.05.2011	&	55707.390	&	13.72	&	14.29	&	14.95	&	15.95	&	Sch	&	FLI	\\
09.06.2011	&	55722.425	&	13.72	&	14.37	&	14.99	&	16.05	&	Sch	&	FLI	\\
21.06.2011	&	55734.420	&	13.81	&	14.42	&	15.06	&	16.07	&	Sch	&	FLI	\\
22.06.2011	&	55735.452	&	13.66	&	14.26	&	14.86	&	15.89	&	Sch	&	FLI	\\
23.06.2011	&	55736.443	&	13.70	&	14.34	&	14.92	&	15.94	&	Sch	&	FLI	\\
24.06.2011	&	55737.443	&	13.77	&	14.34	&	14.95	&	15.95	&	Sch	&	FLI	\\
27.07.2011	&	55770.361	&	13.61	&	14.19	&	14.78	&	15.79	&	Sch	&	FLI	\\
23.08.2011	&	55797.323	&	13.72	&	14.32	&	14.93	&	15.94	&	Sch	&	FLI	\\
24.08.2011	&	55798.301	&	13.76	&	14.38	&	15.01	&	16.03	&	Sch	&	FLI	\\
25.08.2011	&	55799.318	&	13.63	&	14.21	&	14.82	&	15.79	&	Sch	&	FLI	\\
23.09.2011	&	55828.249	&	13.73	&	14.29	&	14.91	&	15.89	&	Sch	&	FLI	\\
27.11.2011	&	55893.176	&	13.68	&	14.27	&	14.88	&	15.94	&	Sch	&	FLI	\\
28.11.2011	&	55894.175	&	13.75	&	14.35	&	14.98	&	15.98	&	Sch	&	FLI	\\
29.11.2011	&	55895.167	&	13.71	&	14.27	&	14.88	&	15.88	&	Sch	&	FLI	\\
30.11.2011	&	55896.179	&	13.71	&	14.30	&	14.92	&	15.95	&	Sch	&	FLI	\\
29.12.2011	&	55925.169	&	13.69	&	14.31	&	14.92	&	15.94	&	Sch	&	FLI	\\
01.01.2012	&	55928.185	&	13.74	&	14.34	&	14.97	&	16.01	&	Sch	&	FLI	\\
16.03.2012	&	56003.563	&	13.69	&	14.26	&	14.89	&	15.92	&	Sch	&	FLI	\\
13.04.2012	&	56030.508	&	13.83	&	14.46	&	15.09	&	16.12	&	Sch	&	FLI	\\
12.05.2012	&	56060.415	&	13.79	&	14.36	&	14.99	&	15.99	&	Sch	&	FLI	\\
20.05.2012	&	56068.399	&	13.82	&	14.45	&	15.06	&	15.94	&	Sch	&	FLI	\\
12.06.2012	&	56091.398	&	13.74	&	14.36	&	14.99	&	16.02	&	Sch	&	FLI	\\
13.06.2012	&	56092.383	&	13.76	&	14.38	&	14.98	&	16.01	&	Sch	&	FLI	\\
17.06.2012	&	56096.398	&	13.63	&	14.17	&	14.77	&	15.76	&	Sch	&	FLI	\\
11.07.2012	&	56120.377	&	13.69	&	14.29	&	14.93	&	15.97	&	Sch	&	FLI	\\
12.07.2012	&	56121.338	&	13.79	&	14.40	&	15.01	&	15.99	&	Sch	&	FLI	\\
13.07.2012	&	56122.399	&	13.67	&	14.23	&	14.83	&	15.81	&	Sch	&	FLI	\\
14.07.2012	&	56123.387	&	13.74	&	14.38	&	14.99	&	16.02	&	Sch	&	FLI	\\
19.08.2012	&	56159.351	&	13.70	&	14.30	&	14.93	&	15.90	&	Sch	&	FLI	\\
20.08.2012	&	56160.329	&	13.69	&	14.25	&	14.88	&	15.88	&	Sch	&	FLI	\\
21.08.2012	&	56161.346	&	13.75	&	14.37	&	15.03	&	16.02	&	Sch	&	FLI	\\
22.08.2012	&	56162.338	&	13.73	&	14.32	&	14.94	&	15.94	&	Sch	&	FLI	\\
04.09.2012	&	56175.426	&	13.77	&	14.39	&	15.01	&	16.05	&	Sch	&	FLI	\\
05.09.2012	&	56176.272	&	13.79	&	14.38	&	15.03	&	16.02	&	Sch	&	FLI	\\
08.09.2012	&	56179.272	&	13.79	&	14.41	&	14.99	&	16.01	&	60-cm	&	FLI	\\
09.09.2012	&	56180.364	&	13.72	&	14.25	&	14.84	&	15.85	&	60-cm	&	FLI	\\
22.09.2012	&	56193.339	&	13.76	&	14.38	&	15.01	&	16.01	&	Sch	&	FLI	\\
23.09.2012	&	56194.322	&	13.70	&	14.30	&	14.89	&	15.90	&	Sch	&	FLI	\\
07.10.2012	&	56208.210	&	13.72	&	14.32	&	14.95	&	15.99	&	Sch	&	FLI	\\
08.10.2012	&	56209.204	&	13.70	&	14.29	&	14.91	&	15.91	&	Sch	&	FLI	\\
09.10.2012	&	56210.193	&	13.70	&	14.27	&	14.87	&	15.86	&	Sch	&	FLI	\\
10.10.2012	&	56211.461	&	13.73	&	-	&	14.98	&	-	&	Sch	&	FLI	\\
25.10.2012	&	56226.240	&	13.72	&	14.32	&	14.97	&	15.93	&	Sch	&	FLI	\\
26.10.2012	&	56227.401	&	13.72	&	14.34	&	14.97	&	15.93	&	Sch	&	FLI	\\
30.10.2012	&	56231.322	&	13.79	&	14.40	&	15.01	&	16.05	&	60-cm	&	FLI	\\
17.11.2012	&	56249.172	&	13.77	&	14.38	&	14.99	&	16.02	&	Sch	&	FLI	\\
18.11.2012	&	56250.181	&	13.70	&	14.28	&	14.91	&	15.91	&	Sch	&	FLI	\\
31.12.2012	&	56293.254	&	13.72	&	14.32	&	14.92	&	15.94	&	Sch	&	FLI	\\
01.01.2013	&	56294.228	&	13.67	&	14.23	&	14.83	&	15.81	&	60-cm	&	FLI	\\
03.01.2013	&	56296.273	&	13.74	&	14.30	&	14.87	&	15.85	&	60-cm	&	FLI	\\
16.01.2013	&	56309.243	&	13.73	&	14.32	&	14.92	&	15.96	&	Sch	&	FLI	\\
05.02.2013	&	56329.196	&	13.62	&	14.21	&	14.80	&	15.83	&	Sch	&	FLI	\\
06.03.2013	&	56357.595	&	13.77	&	14.32	&	14.93	&	-	&	60-cm	&	FLI	\\
10.04.2013	&	56392.510	&	13.74	&	14.37	&	15.03	&	15.98	&	Sch	&	FLI	\\
11.04.2013	&	56394.482	&	13.66	&	14.23	&	14.86	&	15.86	&	Sch	&	FLI	\\
02.05.2013	&	56415.418	&	13.71	&	14.31	&	14.91	&	15.93	&	Sch	&	FLI	\\
15.05.2013	&	56428.404	&	13.68	&	14.26	&	14.87	&	15.85	&	60-cm	&	FLI	\\
17.05.2013	&	56430.408	&	13.80	&	14.39	&	15.02	&	16.03	&	60-cm	&	FLI	\\
30.05.2013	&	56443.367	&	13.69	&	14.26	&	14.86	&	15.84	&	Sch	&	FLI	\\
31.05.2013	&	56444.354	&	13.61	&	14.20	&	14.82	&	15.81	&	Sch	&	FLI	\\
04.08.2013	&	56509.288	&	13.79	&	14.40	&	15.05	&	16.07	&	Sch	&	FLI	\\
05.08.2013	&	56510.369	&	13.62	&	14.17	&	14.77	&	15.75	&	60-cm	&	FLI	\\
05.08.2013	&	56510.386	&	13.61	&	14.17	&	14.80	&	15.78	&	Sch	&	FLI	\\
06.08.2013	&	56511.411	&	13.79	&	14.43	&	15.06	&	16.08	&	Sch	&	FLI	\\
06.08.2013	&	56511.413	&	13.77	&	14.38	&	15.02	&	16.03	&	60-cm	&	FLI	\\
07.08.2013	&	56512.398	&	13.68	&	14.26	&	14.90	&	15.89	&	Sch	&	FLI	\\
07.08.2013	&	56512.403	&	13.66	&	14.20	&	14.80	&	15.81	&	60-cm	&	FLI	\\
08.08.2013	&	56513.382	&	13.64	&	14.19	&	14.78	&	15.83	&	60-cm	&	FLI	\\
09.09.2013	&	56514.350	&	13.78	&	14.42	&	15.07	&	16.05	&	60-cm	&	FLI	\\
12.08.2013	&	56517.284	&	13.70	&	14.29	&	14.89	&	15.85	&	60-cm	&	FLI	\\
04.09.2013	&	56540.274	&	13.66	&	14.26	&	14.86	&	15.91	&	Sch	&	FLI	\\
05.09.2013	&	56541.322	&	13.73	&	14.32	&	14.95	&	15.96	&	Sch	&	FLI	\\
06.09.2013	&	56542.381	&	13.62	&	14.21	&	14.79	&	15.81	&	Sch	&	FLI	\\
11.09.2013	&	56547.383	&	13.67	&	14.23	&	14.84	&	15.80	&	60-cm	&	FLI	\\
14.09.2013	&	56550.363	&	13.70	&	14.23	&	14.86	&	15.89	&	60-cm	&	FLI	\\
07.11.2013	&	56604.264	&	13.71	&	14.30	&	14.93	&	15.95	&	60-cm	&	FLI	\\
28.12.2013	&	56655.200	&	13.70	&	14.28	&	14.88	&	15.88	&	Sch	&	FLI	\\
29.12.2013	&	56656.180	&	13.64	&	14.21	&	14.81	&	15.82	&	Sch	&	FLI	\\
30.12.2013	&	56657.193	&	13.84	&	14.48	&	15.11	&	16.13	&	Sch	&	FLI	\\
23.01.2014	&	56681.189	&	13.73	&	14.30	&	14.91	&	15.91	&	Sch	&	FLI	\\
22.03.2014	&	56738.590	&	13.85	&	14.48	&	15.10	&	16.13	&	Sch	&	FLI	\\
21.05.2014	&	56799.425	&	13.71	&	14.29	&	14.89	&	15.91	&	Sch	&	FLI	\\
28.06.2014	&	56837.417	&	13.73	&	14.36	&	15.00	&	16.04	&	Sch	&	FLI	\\
29.06.2014	&	56838.398	&	13.76	&	14.38	&	15.01	&	16.03	&	Sch	&	FLI	\\
24.07.2014	&	56863.339	&	13.70	&	14.27	&	14.87	&	15.90	&	Sch	&	FLI	\\
25.07.2014	&	56864.356	&	13.82	&	14.43	&	-	&	-	&	Sch	&	FLI	\\
03.08.2014	&	56873.386	&	13.73	&	14.35	&	14.95	&	15.98	&	Sch	&	FLI	\\
04.08.2014	&	56874.407	&	13.70	&	14.25	&	14.86	&	15.82	&	Sch	&	FLI	\\
18.08.2014	&	56888.450	&	13.73	&	14.32	&	14.93	&	15.94	&	Sch	&	FLI	\\
19.08.2014	&	56889.365	&	13.67	&	14.27	&	14.88	&	15.88	&	Sch	&	FLI	\\
13.12.2014	&	57005.181	&	13.68	&	14.30	&	14.84	&	15.83	&	Sch	&	FLI	\\
14.12.2014	&	57006.220	&	13.66	&	14.25	&	-	&	-	&	Sch	&	FLI	\\
\hline \hline
\end{longtable}}
\twocolumn
\twocolumn

\end{appendix}

\end{document}